\documentclass[a4paper,11pt]{article}

\usepackage[a4paper,
  left=2.5cm, right=2.5cm,
  top= 3cm, bottom=4cm]{geometry}
\usepackage{amsmath}
\usepackage{oldgerm}
\usepackage{amssymb}
\usepackage{bbm}
\usepackage[dvips]{graphicx}
\usepackage{epsfig}
\usepackage{color} 
\usepackage{cite}
\usepackage{epic}
\usepackage{hyperref}

\usepackage{empheq}

\numberwithin{equation}{section}

\usepackage[utf8]{inputenc}

\usepackage{graphicx}
\usepackage{mathrsfs}
\usepackage{amsmath}
\usepackage{amssymb}
\usepackage{braket} 
\usepackage{hyperref}
\usepackage{frontespizio}
\usepackage{comment}
\usepackage{epsfig}
\usepackage{float}
\usepackage{color}
\usepackage[title]{appendix}
\usepackage{multirow}
\usepackage{tikz}
\usetikzlibrary{decorations.pathmorphing}
\usepackage{varwidth}
\usepackage{tikz}
\usetikzlibrary{backgrounds,patterns,calc}
\usepackage{xparse}
\usepackage{caption}
\usepackage{enumitem}
\setlist[itemize]{leftmargin=*}

\usetikzlibrary{decorations.pathmorphing}

\tikzset{zigzag/.style={decorate, decoration=zigzag}}

\newcommand{\citep}{\cite}

\newcommand{\bea}{\begin{eqnarray}}
\newcommand{\eea}{\end{eqnarray}}
\newcommand{\be}{\begin{equation}}
\newcommand{\ee}{\end{equation}}
\newcommand{\ba}{\begin{align}}
\newcommand{\ea}{\end{align}}

\usepackage{subfigure}

\def\0{{\boldsymbol 0}}

\def\m{{\vec{m}}}

\def\S{\mathcal S}
\def\M{\mathcal M}

\title{}
\author{}

\numberwithin{equation}{section}

\begin{document}
%
%
\begin{titlepage}

\vspace{-0.5cm} {\flushright {\small{.}}} \\

%
\begin{center}

\vspace{1.2truecm}

{\huge\bf{
Quantum Transitions,  Detailed Balance, Black Holes and Nothingness\\[0.3cm]}}

\vspace{1.2truecm}

{\fontsize{10.5}{18}\selectfont
{\bf Sebasti\'an C\'espedes,${}^{\rm a}$ Senarath de Alwis,${}^{\rm b}$ Francesco Muia ${}^{\rm c}$ and Fernando Quevedo${}^{\rm c, d}$
}}
\vspace{.5truecm}

{\small{\it $^{a}$ Department of Physics, Imperial College London,
London, SW7 2AZ, UK}}\\
  {\small {\it $^{b}$ Physics Department, University of Colorado, Boulder, CO 80309 USA}}\\
{\small {\it $^{c}$ DAMTP,  Centre for Mathematical Sciences,  University of Cambridge,\\ Wilberforce Road,  Cambridge, CB3 0WA, UK}}\\
{\small {\it $^{d}$ Perimeter Institute for Theoretical Physics
31 Caroline Street North,\\ Waterloo ON, Canada.}}
 
  \vskip 2.2cm

 \begin{abstract}
{\small{{We consider vacuum transitions  by bubble nucleation among 4D vacua  with different values and signs of the cosmological constant $\Lambda $, including both up and down tunnelings. Following the Hamiltonian formalism, we explicitly compute the decay rates for all possible combinations of initial and final values of $\Lambda $ and find that up-tunneling is allowed starting not only from   dS spacetime but also from  AdS and Minkowski spacetimes. We trace the difference with the Euclidean approach, where these transitions are found to be forbidden, to the difference of treating the latter spacetimes as pure (vacuum) states rather than mixed states with correspondingly vanishing or infinite entropy. We point out that these transitions are best understood as limits of the corresponding transitions with black holes  in the zero mass limit $M\rightarrow 0$.  We 
find that detailed balance is satisfied provided we use the Hartle-Hawking sign of the wave function for nucleating space-times. In the formal limit $\Lambda \rightarrow -\infty $, the transition rates for AdS to dS agree with both the  Hartle-Hawking and Vilenkin amplitudes for the creation of dS from nothing. This is consistent with a proposal of Brown and Dahlen to define  `nothing'  as AdS in this limit. For $M\neq 0$ detailed balance is satisfied only in a range of mass values.  We compute the bubble trajectory after nucleation and find that, contrary to the   $M=0$ case, the trajectory  does  not  correspond to the open universe slicing of dS. We briefly discuss the relevance of our results to the string landscape.
 
 }}}
 \end{abstract}
\end{center}

\end{titlepage}






\tableofcontents

\renewcommand*{\thefootnote}{\arabic{footnote}}
\setcounter{footnote}{0}

\newpage

\section{Introduction}

Understanding the vacuum state of our universe is of paramount importance. The vacuum energy determines in great part the geometry of spacetime once gravity is considered. Quantum transitions between states of different vacuum energies should provide fundamental information on how to understand the vacuum states in a fully-fledged theory of quantum gravity. In particular it should help us to better understand the string theory landscape.\footnote{See~\cite{Cicoli:2023opf} for a recent review of string cosmology including attempts to  obtain de Sitter space in string theory. For alternatives such as quintessence see for instance \cite{Cicoli:2018kdo, Hebecker:2019csg, Cicoli:2021fsd}.} 
\vskip 1mm

Most of the progress in this direction has been made by using  semiclassical techniques.
In particular, the Euclidean approach pioneered by Coleman and collaborators \cite{Coleman:1980aw} borrows results from non-gravitational field theories  \cite{Coleman:1977py, Callan:1977pt} and makes a concrete proposal for the transition rate per unit volume of such a quantum transition. For the simplest set-up of a scalar field theory with a potential energy with two minima $A$ and $B$ with different values of the vacuum energy $V_A > V_B$  the Coleman-De Luccia (CDL) prescription~\cite{Coleman:1980aw} for the transition rate is:
\be
\Gamma_{A\rightarrow B} \propto e^{-\left[S({\rm bounce})- S(A)\right]/\hbar} \,, \label{bounce}
\ee
where $ S({\rm bounce})$ refers to the Euclidean action evaluated at the instanton solution (bounce) that mediates between the two vacua, and $S(A)$ is the Euclidean action evaluated at the background spacetime $A$. The process amounts to the creation of a bubble of vacuum $B$ in the background of vacuum $A$ mediated by the bounce. The explicit calculation boils down to determining the bounce as a solution of the Euclidean equations and plugging it into Eq.~\eqref{bounce}. We emphasise that Eq.~\eqref{bounce} is only a proposal and lacks an explicit derivation.
\vskip 1mm

In principle the same instanton can mediate the up-tunneling transition, from $B$ to $A$ ~\cite{Lee:1987qc},
\be
\Gamma_{B\rightarrow A} \propto e^{-\left[S({\rm bounce})- S(B)\right]/\hbar} \,, \label{bounce2}
\ee
and cancels in  the ratio of transitions
 \be
 \frac{\Gamma_{B\rightarrow A} }{\Gamma_{A\rightarrow B} } \propto e^{\left[S({B})- S(A)\right]/\hbar} \,.
 \ee
 However, it has been argued that only in the case in which both vacuum energies are positive (de Sitter to de Sitter i.e. dS $\rightarrow$ dS transition), is up-tunneling  allowed since otherwise the background contribution $S(B)$ is negative and infinite, implying $\Gamma_{B\rightarrow A} \rightarrow 0$~\cite{Lee:1987qc}. In the allowed case 
  \be
 \frac{\Gamma_{B\rightarrow A} }{\Gamma_{A\rightarrow B} } \propto e^{\left[S({B})- S(A)\right]/\hbar} \propto e^{-\left[\S_B- \S_A\right]/\hbar} \,, \label{detailed}
 \ee
 where $\S_A=\pi/H_A^2, \, \S_B=\pi/H_B^2$ are the entropies of the two de Sitter spaces and $H^2=\Lambda/3$ determines the horizon radius $R_{dS}=1/H$ for a dS space~\cite{Gibbons:1976ue}. Eq. \eqref{detailed} is the statement of detailed balance: the ratio of probabilities is determined by the corresponding  number of available states which is measured by the exponential of the entropies. 
 \vskip 1mm
 
From this perspective, the fact that there are no up-tunneling vacuum transitions starting from Minkowski $\M$ or anti de Sitter (AdS) spacetimes corresponds to the fact that, contrary to the case of de Sitter space,  the volumes of both AdS and $\M$ are infinite. This implies that the corresponding background actions $S(B)$ diverge $S(B)\rightarrow -\infty$ and the transition rate vanishes.\footnote{Note that assuming that the partition function is approximated by one dS space-time, the Euclidean action is $S_E=\beta F=\beta E-{\S}$. If $E=0$ then $S_E=-\S$.} 
\vskip 1mm

Note that, following the dS analogy, this is consistent with assigning an infinite entropy to both $\M$ and AdS.\footnote{It should be noted however that assigning infinite Euclidean action to Minkowski and AdS is a consequence of ignoring the infrared regulator that Gibbons and Hawking \cite{Gibbons:1976ue},\cite{Gibbons:1977mu}  included in their calculation of black hole entropy. To get the well-known formula for black hole entropy (in Minkowski (M) or AdS space) one needs this regulator  which effectively implies that the Euclidean action and hence the entropy of empty  M or AdS space is zero. For a recent review of these entropy calculations as well as a Hamiltonian version which gives the same result see~\cite{deAlwis:2023tab}.} However, Eq.~\eqref{bounce} is only the result of an educated guess and it is not derived from a well defined prescription in quantum  gravity\footnote{Also the assumption that the bounce solution in gravity inherits the $O(4)$ symmetry that was derived  for flat space field theory is not justified.}, therefore we should be open to the possibility that it may not fully capture the relevant physics. 
\vskip 1mm

Note also that, contrary to dS, Minkowski and AdS do not have horizons and therefore assigning an entropy in terms of an area is not an option. On counting states, we know that the Hilbert space $\mathcal H$ in both cases is infinite which could indicate an infinite entropy. However, as  vacuum states both cases would be pure states with vanishing entropies.\footnote{In principle we can assign vanishing entropy to a full dS spacetime also, however in practice we are interested in the regions that are accessible to a given observer which has no access to the region beyond the horizon. Tracing out the states beyond the horizon gives rise to a non-vanishing entropy.} 
Therefore, thinking in terms of entropies we may ask two different questions: what is the transition rate between maximally mixed states in the Hilbert space or what is the transition rate only from the vacuum state (a  pure state). In the latter, the entropy vanishes and the corresponding up-tunneling amplitude is allowed. We will see that this is what is obtained once we use the Hamiltonian approach rather than the Euclidean approach to compute vacuum transitions.
 \vskip 1mm
 
 In the Hamiltonian approach developed by Fischler, Morgan and Polchinski (FMP)\footnote{The FMP calculation was an attempt to understand from first principles in the Hamiltonian framework the singular instanton calculation of Farhi, Guth and Guven (FGG) \cite{Farhi:1989yr} whose validity was questioned even  by the original authors. Further concerns about the validity of both FMP and FGG have been raised over the years \cite{Freivogel:2005qh, Fu:2019oyc, Susskind:2021yvs}. See \cite{DeAlwis:2019rxg} for addressing the concerns in \cite{Freivogel:2005qh}. We will address the points raised by \cite{Susskind:2021yvs} in Sec.~\eqref{sec:susskind}. } \cite{Fischler:1989se,Fischler:1990pk} (see also \cite{Bachlechner:2016mtp, DeAlwis:2019rxg}),
vacuum transitions are defined as generalisations of the standard WKB method in Quantum Mechanics. This method has the advantage that the transition rate is actually computed (not guessed) following the standard WKB approach. No Wick rotations to Euclidean spaces are needed, nor any assumption about the dominant instanton contribution needs to be made. 
\vskip 1mm

On the other hand, the Hamiltonian approach lacks the flexibility of the Euclidean approach. For example the more general set-up in terms of a potential with two minima,  studied by CDL,  has not yet  been addressed in this formalism.\footnote{It should be borne in mind though that in practice the CDL procedure has proved amenable to calculation only in the thin wall regime - which is in effect equivalent to the brane nucleation calculation of the Hamiltonian approach and indeed to the Euclidean version of the brane nucleation calculation of Brown and Teitelboim \cite{Brown:1988kg}.} So far,  the Hamiltonian approach,  has only been developed for the simplest set-up of two  spacetimes of different cosmological constants $\Lambda_A, \Lambda_B$ separated by a wall of tension $\kappa$ that determines the bubble.  
\vskip 1mm

In any case, this is enough for our purposes in this paper. Precisely this simple set-up is what allows the calculations to be explicit and reliable. This is because after solving the energy and momentum constraints the problem reduces to a quantum mechanics problem with a single degree of freedom: the location of the wall $\hat R$. The resulting equation of motion is simple: $\dot{\hat R}^2 +V(\hat R)=-1$ where $V(\hat R)$ a calculable function of $\hat R$ determined from the parameters of the metric and the wall tension \cite {Blau:1986cw}. Then the standard quantum mechanics rules apply in terms of a transition through a barrier for a potential energy $V(\hat R)$.
\vskip 1mm

\begin{figure}[h!] 
\begin{center} 
\includegraphics[scale=0.5, trim = 1cm 2cm 0cm 5cm, clip]{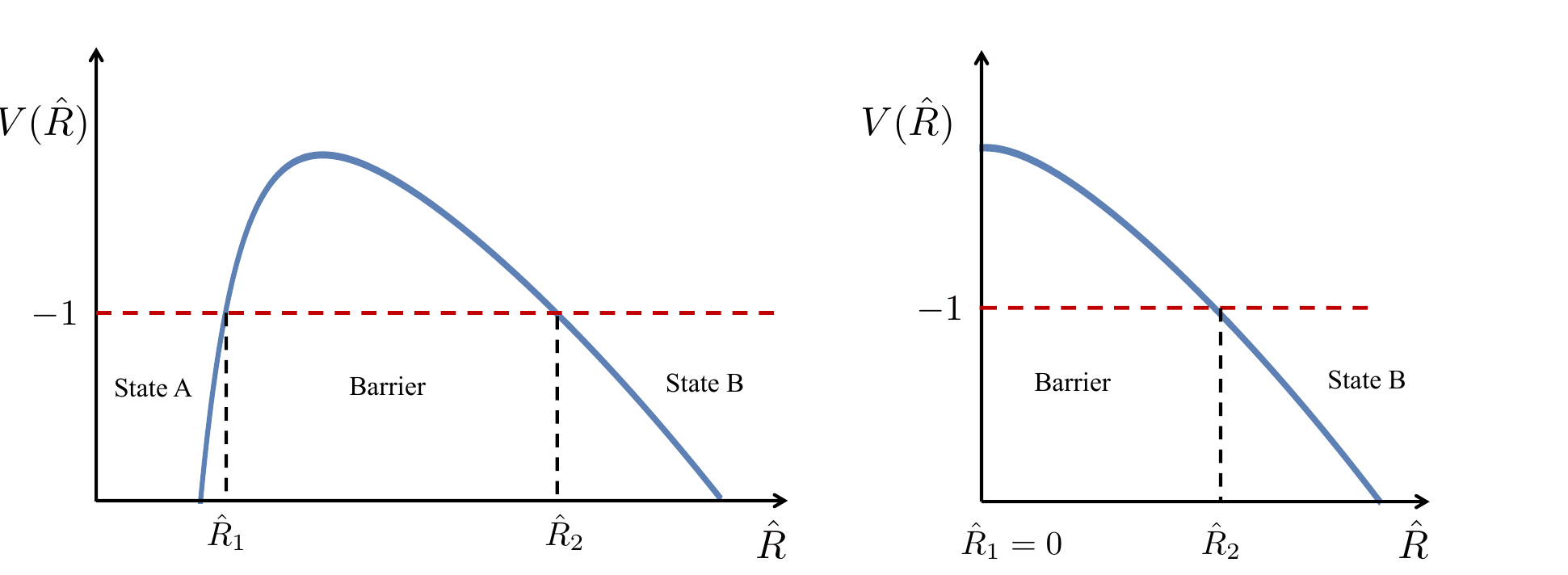}
\caption{\footnotesize{Two realisations of the potential for the bubble wall $\hat R$. On the left, a bubble is materialised in region $A$ and grows until it reaches the turning point $\hat R_1$ and classically bounces but quantum mechanically can tunnel to region $B$ at the second turning point to continue expanding. The  WKB approximation can be used in all the regions of the potential outside and inside the barrier. This is a typical situation for black hole geometries. On the right there is only one turning point (the first turning point has moved to zero) and the bubble materialises directly in state $B$. This is a typical potential for pure dS or AdS that can be obtained by setting the black hole mass to zero from the black hole geometry.} \label{fig:potentialRhat}}
\end{center} 
\end{figure}

In this approach the transition probability from a state $A$ to $B$ is defined as the ratio of the two squares of the corresponding wave functions:
 
 \be
 \Gamma_{A\rightarrow B}=\frac{|\Psi(\hat R_2)|^2}{|\Psi(\hat R_1)|^2} \,,
 \ee
where $\hat R_1$ and $\hat R_2$ are the two turning points that are determined from the potential barrier $V(\hat R)$, see the left panel of Fig.~\ref{fig:potentialRhat}. The classical picture is that the brane, which contains inside it spacetime B, is nucleated with $\hat R = 0$ in spacetime $A$, grows to a radius $\hat{R}_1$, is reflected back and collapses back to zero. There is also a classical trajectory where the brane comes in from infinity with infinite size (at least in a non-compact space such as a black hole space) hits the barrier at $\hat{R}_2$ and is reflected back. Quantum mechanically the brane can tunnel between these two classical configurations. The wave functions can be written in terms of the WKB approximation as in standard Quantum Mechanics
\be
\Psi(\hat R)=a e^I +b e^{-I} \,,
\ee
where $I=iS$ is $i$ times the action evaluated at  $\hat R$. For the case when one of the two exponentials dominate, the transition rate can be written in terms of a difference between two actions, similar to Eq.~\eqref{bounce} but as we will see they differ in important ways.

In the next sections we will provide a short summary of this prescription and its application to explicitly compute the transition rates between vacuum states,  we will also include transitions between spacetimes with black holes and in the case that the true vacuum is de Sitter we will study its geometry. 
\vskip 1mm

Let us summarise the main results of this article:

\begin{itemize}
\item We explicitly compute the  rates for transitions between any of  dS and AdS states including both up and down tunneling and provide explicit expressions for each of the transition rates. The cases corresponding to up-tunneling from AdS are new results whereas the others are known and wherever the non-vanishing result is known we  agree with the previous results in the literature.

\item We consider Minkowski spacetime $\M$ in three different limits. First, starting from pure dS with curvature $\Lambda>0$  and taking the limit $\Lambda\rightarrow 0$: in this case we obtain vanishing up-tunneling transition as in the Euclidean case. Second, we start with an AdS spacetime with curvature $\Lambda<0$ and take the limit $\Lambda \rightarrow 0$. In this case we get a finite transition amplitude. We interpret the results by noticing that in the dS limit case, the entropy $\S\propto 1/\sqrt{\Lambda}\rightarrow \infty$ whereas in the AdS limit case the background contribution vanishes. This corresponds  to a vanishing entropy for AdS which is inherited in the Minkowski limit. The third limit is the one taken by FMP corresponding to the $M\rightarrow 0$ of  Schwarzschild spacetime. This coincides with the AdS limit and gives the same finite transition rate.

\item Having computed up-tunneling transitions from AdS we also compute the limit $\Lambda \rightarrow -\infty$. In this case up-tunneling to dS gives the well-known Vilenkin \footnote{Also known as tunneling or Vilenkin-Linde wave function.} and Hartle-Hawking~\cite{Vilenkin:1982de, Hartle:1983ai, Vilenkin:1984wp, Linde:1983mx, Rubakov:1984bh} transitions from nothing to dS. This $\Lambda \rightarrow -\infty$ limit is consistent with the proposal of Brown and Dahlen~\cite{Brown:2011gt} for a field theoretic definition of `nothing' precisely as AdS with infinite curvature, motivated by an interpretation of the bubble of nothing transition\footnote{However, they argued that the up-tunneling transition is forbidden and questioned the validity of the Hartle-Hawking and Vilenkin set-ups. Our conclusion in this work  is that  not only are these transitions  allowed but that they reproduce exactly the same results as Hartle-Hawking and Vilenkin wave functions.}. Therefore we find consistency between the two definitions of nothing (the decay to the bubble of nothing and the creation of dS from nothing). Furthermore we also explicitly compute transitions from nothing to AdS and $\M$ which were previously thought not to be allowed. 

\item A non-trivial check of our results is that we obtain detailed balance in all the transitions as long as $M\rightarrow 0$.

\item A further non-trivial check of our results is to generalise the transitions to include mass $M$ black hole backgrounds, which are the most general solutions with spherical symmetry which is the symmetry of the bubble/wall system. In this case we can compute the transitions and reproduce our previous results in the limit of $M\rightarrow 0$. 

\item In general, the black hole transitions depend on three regions for the values of $M$. Generalising FMP we compute the bulk contribution to the transitions in all regimes. Contrary to the $M=0$ limit some transitions are not allowed. However the Schwarzschild de Sitter (SdS) transition SdS$\rightarrow$ SdS is allowed and reproduces previous proposals. In particular it is consistent with detailed balance for small black holes (to be defined below). A point that has been recently questioned in \cite{Susskind:2021yvs}. A detailed discussion of the issues raised in this reference is given in Sec.~\ref{sec:susskind}.

\item When the true vacuum is de Sitter we compute the trajectory of the wall.  We will see that in the  case when $M\neq 0$,  contrary to that when $M=0$, we find that it does not follow a geodesic which favors  open universe slicing.

\end{itemize}

The order of the presentation is as expressed in the table of contents and may not need further explanation. The SdS $\rightarrow$ SdS case is technically more complicated due to the different horizons. It is done in full detail and in order to facilitate the readership we move it to an appendix that includes all technical details. 

In general, our results imply that populating  the  string landscape is straightforward and provide some hints on how this population may happen. In particular it incorporates the creation from nothing on the same footing as the other transitions.

\section{Hamiltonian Approach to Vacuum Transitions}

Let us start reviewing vacuum transitions from the Hamiltonian approach as initiated by Fischler, Morgan and Polchinski (FMP) \cite{Fischler:1990pk}.  Starting with the spherically symmetric metric
 \be
 ds^2=-N_t^2 dt^2+L(r,t)^2 (dr+N_r dt)^2+R(r,t)^2 d\Omega_2^2 \,,
 \ee
in order to address the vacuum transition problem FMP considered the bulk-brane system with the brane (or wall) at $r=\hat r$ separating two regions with different cosmological constants $\Lambda_\pm$
and the following action:
\be
S=S_{bulk}+S_{brane}-\int d^4x \sqrt{-g}\left(\Lambda_+\Theta(r-\hat{r})+\Lambda_-\Theta(\hat{r}-r)\right) \,,
\ee
with standard Einstein-Hilbert   $S_{bulk}$ and brane action $S_{brane} $ respectively  and
 with $\Theta$ the step function. 

FMP reduced the vacuum transition problem to solving for the quantum mechanics of the
brane (assumed spherically symmetric) with a wave function $\Psi(\hat{R})$
which solves the Wheeler-deWitt  equation. In the leading WKB approximation this
implies solving the momentum and Hamiltonian constraints while satisfying
the matching conditions at the brane~\cite{Blau:1986cw}.

\be
\label{eq:JunctionConditions}
\frac{R'(\hat{r}\pm\epsilon)}{\hat{L}}=\frac{1}{2\kappa\hat{R}}\left(\hat{A}_{{\rm I}}-\hat{A}_{{\rm O}}\right)\mp\frac{\kappa}{2}\hat{R} \,.
\ee
 
Here $\kappa=4\pi G\sigma$  where $\sigma$ is the tension of the wall and $\hat{A}_{I,O}$ are the static metric functions evaluated at $r=\hat r$. The indices $I,O$ refer to interior and exterior of the wall;
 
 \be
 A_{\alpha}  = 1-\frac{2GM_{\alpha}}{R}\mp H_{\alpha}^{2}R^{2}\,,  \qquad \alpha=I,O \,,
\ee
 where $ \pm H_{\alpha}^{2}=\frac{8\pi G}{3}\Lambda_{\alpha} $, with upper sign corresponding to de Sitter and the lower one to anti-de Sitter, and $M_\alpha$ is the standard integration constant corresponding to a black hole mass.
 Using the explicit expression for $\pi_{L}$ (in the gauge $N_{r}=0$)
one can write the (first integral of the) equation of motion for the
brane:
\begin{eqnarray}
\dot{\hat{R}}^{2}+V & = & -1 \,,\label{eq:EofM}\\
V & = & -\frac{1}{(2\kappa\hat{R})^{2}}\left((\hat{A}_{{\rm I}}-\hat{A}_{{\rm O}})-\kappa^{2}\hat{R}^{2}\right)^{2}+(\hat{A}_{{\rm O}}-1)\label{eq:V}\\
 & = & -\frac{1}{(2\kappa\hat{R})^{2}}\left((\hat{A}_{{\rm I}}-\hat{A}_{{\rm O}})+\kappa^{2}\hat{R}^{2}\right)^{2}+(\hat{A}_{{\rm I}}-1),\label{eq:Valt}
\end{eqnarray}
which is unbounded from below with a local maximum and one or two turning points at $V=-1$ depending on the parameters $H_\alpha, M_\alpha$. 
So the equation $V=-1$ gives the (two) turning points for $\hat{R}$,
$R_{1}<R_{2}$ for the classical motion of the brane. At a turning
point we see from Eq.~\eqref{eq:V} that $\hat{A}_{{\rm O}}>0$ and from Eq.~\eqref{eq:Valt} that $\hat{A}_{{\rm I}}>0$. 

The classical turning points of the geometry correspond to vanishing conjugate momenta for $L$ ($\pi_L=0$ which also implies $\pi_R=0$) \cite{Fischler:1989se}, which gives:
\be
\label{eq:TurningPoints}
\frac{R'^{2}}{L^{2}} = A(R) = 1-\frac{2MG}{R}\mp H^{2}R^{2} \,.
\ee
For $r=\hat r$ these correspond to the turning points of the potential (i.e. the solutions of $V=-1$).

The sign of $R'(\hat r)$ plays an important role since $R'$ is proportional to the extrinsic curvature $\hat K$ and indicates if the wall at $r=\hat r$ is bent towards the interior or exterior regions.

The transition probability\footnote{Note that the Wheeler-DeWitt equation is like a time-independent Schrodinger equation in that there is no notion of time evolution. Consequently there is no obvious way to discuss the rate/lifetime of a state. In the case of non-gravitational field theory too, in order to discuss a rate one needs to construct a wave packet. For a discussion of this and for a critique of the standard Coleman approach to tunneling in field theory see \cite{Andreassen:2016cvx} and \cite{deAlwis:2023gth}. } from the initial state $\mathcal{B}$ to the nucleated state $\mathcal{N}$ (including the two spacetimes and the wall) can be written as:
\be
{\mathcal P}\left(\mathcal{B}\rightarrow \mathcal{N}\right)=\frac{\|\Psi(\mathcal{N})\|^2}{\|\Psi(\mathcal{B})\|^2} \,,
\ee
where the numerator can be identified as the squared wave function at the turning point $R_2$ and the denominator with that at $R_1$, as in the previous section.
These can be written in the WKB approximation as

\be
\Psi=a e^I + b e^{-I}, \label{eq:PsiR0}
\ee
 where $I=iS$. For the numerator we will have a bulk contribution $I_{\rm B}$ and a boundary contribution $I_{\rm W}$ which take the form:

\begin{align}
\label{eq:SclassicalBulk}
I_{{\rm B}} & = \frac{\eta}{G}\int_{0}^{\hat{r}-\epsilon}drR\left[ \sqrt{A_{{\rm I}}L^{2}-R'^{2}}-R'\cos^{-1}\left(\frac{R'}{L\sqrt{A_{{\rm I}}}}\right)\right] + \int_{\hat{r}+\epsilon}^{r_{\rm max}}dr \,\left[{\rm I}\rightarrow{\rm O}\right] \,, \\
\label{eq:SclassicalBoundary}
I_{{\rm W}} & = \frac{\eta}{G}\int\delta\hat{R} \, \hat{R}\cos^{-1}\left(\frac{R'}{L\sqrt{\hat{A}}}\right)\bigg|_{\hat{r}-\epsilon}^{\hat{r}+\epsilon} \,.
\end{align}
Using Eq.~\eqref{eq:JunctionConditions} we can write explicitly:
\begin{align}
\label{eq:Integrals}
I_{\rm W} = &- \frac{\eta}{G}\int dR R \cos^{-1} \left(\frac{\frac{2G}{R}(M_{\rm O} - M_{\rm I}) + R^2(\pm H_{\rm O}^2 \mp H_{\rm I}^2 - \kappa^2)}{2 \kappa R \sqrt{1 - \frac{2GM_{\rm O}}{R} \mp H_{\rm O}^2 R^2}}\right)  \nonumber \\
&+ \frac{\eta}{G}\int dR R \cos^{-1} \left(\frac{\frac{2G}{R}(M_{\rm O} - M_{\rm I}) + R^2 (\pm H_{\rm O}^2 \mp  H_{\rm I}^2 + \kappa^2)}{2 \kappa R \sqrt{1 - \frac{2GM_{\rm I}}{R} \mp H_{\rm I}^2 R^2}}\right) \,.
\end{align}

The total action is given by:
\be
I_{\rm tot} = I_{{\rm B}} + I_{{\rm W }} \,,
\ee

Note that at the turning points $R'^2=L^2 A$ and therefore the term in the expression for $I_{\rm B}$ the square root term vanishes and the argument of $\cos^{-1}$ is $\pm 1$. So the only non-vanishing contribution corresponds to $R'/L=-\sqrt{A} \leq 0$. This is a very strong constraint on the allowed values of the extrinsic curvatures. 
On the turning point geometry $\pi_{L}=0$ the first term vanishes
and the second term is non-zero only when the argument of the inverse
cosine is $-1$. This means that the bulk integral is given by

\begin{align}
S_{{\rm B}} & =\frac{i\eta\pi}{G}\left[\int_{0}^{\hat{r}}dr\frac{dR}{dr}R\theta\left(-R'\right)+\int_{\hat{r}}^{r_{\rm max}}dr\frac{dR}{dr}R\theta\left(-R'\right)\right]\label{eq:Bulk}\\
 & =\frac{i\eta\pi}{2G}\left[\left(\hat{R}^{2}-R^{2}(0)\right)\theta\left(-\hat{R}'_{-}\right)+\left(R_{{\rm }}^{2}(r_{\rm max})-\hat{R}^{2}\right)\theta\left(-\hat{R}'_+\right)\right] \,. \label{eq:Bulk-1}
\end{align}

 Note that the sign of $R'$ in the first line is determined by continuity and the sign of $\hat{R}'_{\pm}$ which is determined by the matching
conditions and fixed by the geometry on either side of $\hat{r}$.
It should  also be noted that if the geometry on either or both
sides has horizons then $R(0)$ may have to be replaced by the solution of $A_{I}=0$ (i.e. the smallest horizon) and $R(R_{\rm max})$ by the solution of $A_{O}=0$ since at a horizon the sign of $R'$ will change and will no longer contribute to the integral.

A further condition is to guarantee that the bubble radius at the turning points is  real. Both conditions play a role in the concrete cases we will study next.

\subsection{dS to dS transitions}

The simplest transitions to study are dS to dS for which both down and up-tunneling are allowed. This correspond to the particular cases in which both black hole masses vanish, $M_I=M_O=0$. In the case of an initial dS (with $H=H_{{\rm O}}$) to a final dS (with
$H=H_{{\rm I}}$) we are in the limit $M_{O,I}\rightarrow 0$ had only one turning point so $R_{1}\rightarrow 0,R_{2}\equiv R_{0}$ with

\be
R_{0}^{2}  = \frac{4\kappa^{2}}{(H_{{\rm O}}^{2}-H_{{\rm I}}^{2})^{2}+2\kappa^{2}(H_{{\rm O}}^{2}+H_{{\rm I}}^{2})+\kappa^{4}} \,.\label{eq:RoHH}  
\ee

Also
\begin{eqnarray}
A_{{\rm O}} & = & 1-H_{{\rm O}}^{2}R^{2},\,\,A_{{\rm I}}=1-H_{{\rm I}}^{2}R^{2},\label{eq:AOIHH}\\
V & = & -\frac{1}{4\kappa^{2}}\hat{R}^{2}\left[(H_{{\rm O}}^{2}-H_{{\rm I}}^{2})^{2}+2\kappa^{2}(H_{{\rm O}}^{2}+H_{{\rm I}}^{2})+\kappa^{4}\right],\label{eq:VHH}
\end{eqnarray}
and there is no initial turning point. Also the matching conditions
are now
\begin{equation}
\frac{\hat{R}'_{\pm}}{L}=\frac{1}{2\kappa\hat{R}}(H_{{\rm O}}^{2}-H_{{\rm I}}^{2}\mp\kappa^{2})\hat{R}^{2}\equiv c_{\pm}\hat{R}\,. \label{eq:R'HH}
\end{equation}
 In this case we were able to evaluate explicitly the wall term as well as the 
 bulk term. For a general value
of $\hat{R}$ the latter  can be directly evaluated from Eq.~\eqref{eq:Bulk} and gives 
\begin{equation}
I_{{\rm B}}\left(\hat{R}\right)\equiv iS_{{\rm Bu}}\left(\hat{R}\right)=\frac{\eta\pi}{2G}\left[(\theta(-\hat{R}'_{+})-\theta(-\hat{R}'_{-})\hat{R}^{2}+\theta(-\hat{R}'_{-})H_{{\rm I}}^{-2}\right]\label{eq:SBHH0}.
\end{equation}

To get the tunneling factor we need to find the difference between the actions evaluated at the two turning points. However in the dS to dS case there is no initial turning point - so in effect it becomes $\hat R=0$. To make these WKB tunneling calculations well defined in the absence of an explicit parametrization as in \cite{DeAlwis:2019rxg}, one should really consider this as the limit of the case with   an initial turning point before which there is a classical region. In other words this should be regarded as coming from the $M\rightarrow 0$ limit of the corresponding case with a black hole (see appendix) where we do have two turning points - $R_1$, $R_2$, with 
$R_{1}<R_{<}<R_{>}<R_{2}$, where $R_<$, $R_>$ are given in Eq.~\eqref{eq:R<>}. In the limit where the black hole masses go to zero $R_{1}<R_{<}\rightarrow 0$. Furthermore (see Fig.~\ref{fig:potentialSdS}) we see that $\hat R'_-$ at $\hat R = R_1$ is positive (and so is  $\hat R'_+$). Hence 
at this point  $\theta(-\hat R'_{\pm})$ are both zero. Hence with this definition of the initial turning point in the limit where the black hole masses are zero we have from Eq.~\eqref{eq:SBHH0}, $I_{{\rm B}}(\hat{R}=R_1\rightarrow 0)=0$. Thus defining the difference of the two turning point actions $I$ as $I(R_2)-I(R_1)\equiv I(R_2-R_1)$
 we have (with $R_2=R_0,~R_1=0$)
 \begin{equation}
I_{{\rm B}}\left({R_0-0}\right)\equiv iS_{{\rm Bu}}\left({R_0}-0\right)=\frac{\eta\pi}{2G}\left[(\theta(-\hat{R}'_{+})-\theta(-\hat{R}'_{-})\hat{R}^{2}+\theta(-\hat{R}'_{-})H_{{\rm I}}^{-2}\right] \,.\label{eq:SBHH1}
\end{equation}

Eq.~\eqref{eq:SBHH0} is however not symmetrical between the outside and the inside of the spherical brane. This symmetry is a property of the full bulk (and wall) integrals (see  Sec.~\eqref{subsec:Reflection-symmetry-of} in the appendix) before imposing any turning point conditions. If we do impose this symmetry even at the turning point geometry (where $R'/L)=\pm \sqrt A$) then the above becomes
 
 \begin{equation}
I_{{\rm B}}\left(\hat{R}\right)\equiv iS_{{\rm Bu}}\left(\hat{R}\right)=\frac{\eta\pi}{2G}\left[(\theta(-\hat{R}'_{+})-\theta(-\hat{R}'_{-})\hat{R}^{2}+\theta(-\hat{R}'_{-})H_{{\rm I}}^{-2}+\theta(\hat{R}'_{+})H_{O}^{-2}\right] \,.\label{eq:SBHH}
\end{equation}
This formula explicitly displays the symmetry under $O\leftrightarrow I$ which implies $\hat{R}_{\pm}\leftrightarrow-\hat{R}_{\mp}$. This was also the result was obtained in \cite{DeAlwis:2019rxg} where an explicit parametrization was used to compute the action at a general point (i.e. neither the geometry nor the brane were at a turning point\footnote{In fact with the explicit parametrization used in that paper putting the geometry at a turning point implied that the brane was also at a turning point. In other words the equation above is valid only at the turning points. This is a consequence of the relation $1-H^2_{I,O}a^2=\frac{1-{\hat R}^2/R_0^2 }{1-c^2_{-,+}{\hat R}^2} $ (with the $c$'s constants), which shows that when the geometry is at a turning point $a=H^{-1} $ then so is the brane, i.e. $\hat R=R_0$. Furthermore it shows that at $\hat R=0$ $a=0$, which is allowed since under the barrier dS space is a Euclidean 4-space.\label{foot:explicit}} and hence it automatically had this symmetry. Since we will not have the luxury of such a parametrization in the subsequent analysis it behoves us to understand the results in terms of general properties of the integrals.

To find the bulk action at 
 the initial point   we have to put  $\hat{R}=0$ in  equation \eqref{eq:SBHH}: However as argued below Eq.~\eqref{eq:SBHH0},  $\theta(-\hat{R}'_{-})\rightarrow0$ 
and $\theta(-\hat{R}'_{+})\rightarrow0$ when $\hat R=R_1\rightarrow 0$.
\be
I_{{\rm B}}\left(\hat{R}=0\right)=\frac{\eta\pi}{2G}H_{O}^{-2} \,,
\ee
 In this case we have 
 \begin{equation}
I_{{\rm B}}\left({R_0-0}\right)\equiv iS_{{\rm B}}\left({R_0}-0\right)=\frac{\eta\pi}{2G}\left[(\theta(-\hat{R}'_{+})-\theta(-\hat{R}'_{-})\hat{R}^{2}+\theta(-\hat{R}'_{-})H_{{\rm I}}^{-2}-\theta(-\hat{R}'_{+})H_{O}^{-2}\right] \,.\label{eq:SBHH'}
\end{equation}

The expressions in Eq.~\eqref{eq:SBHH1} and Eq.~\eqref{eq:SBHH'} differ when $\hat R'_+$ is negative.\footnote{Note however that if $\hat R'_+$ is negative then the last term of Eq.~\eqref{eq:SBHH} is zero and the difference of the turning point actions is the same as before.} As explained in detail in the appendix this difference comes from the fact that the explicit parametrization used in
~\cite{DeAlwis:2019rxg} automatically includes the possibility of the brane being created behind the horizon (of an observer at  $R=0$) in the outside dS space.

Also, following ~\cite{DeAlwis:2019rxg}, the integrals appearing in the expression for $I_{\rm W}$ in Eq.~\eqref{eq:Integrals} can be done analytically which we reproduce here for further use: 
\begin{eqnarray}
I_{\rm W}\Big |_{tp} &=& -\frac{\eta}{G} \int_0^{R_0} \delta \hat R \hat R\left [\cos^{-1}\left(\frac{\hat R'_+}{L\sqrt{\hat A_O}}\right)-\cos^{-1}\left(\frac{\hat R'_-}{L\sqrt{\hat A_I}}\right)\right ]\\ \nonumber
&= & -\frac{\eta \pi}{4G} R_0^2\left[\frac{\epsilon\left(\hat R'_+\right)}{1+|c_+|R_0}-\frac{\epsilon\left(\hat R'_-\right)}{1+|c_-|R_0}+2\left(\theta\left(-\hat R'_+\right)-\theta\left(\hat R'_-\right)\right)\right] \,,
\end{eqnarray}
where $ \epsilon(\hat R^{\prime}_{\pm})$ refer to the sign of $ \hat R^{\prime}_{\pm} $. 

Therefore, for general dS (and AdS) transitions the brane contribution being local is finite and can be computed explicitly:
\begin{eqnarray}\notag
I_{\rm W}(R_0-0) 
 &= &\frac{\eta}{G} \left[\frac{\pi}{2}R_{0}^{2} \left(\Theta(-\hat R^{\prime}_{-})-\Theta(-\hat R^{\prime}_{+})\right)+\frac{\pi}{4H_{\rm I}^{2}}\ \epsilon(\hat R^{\prime}_{-})-\frac{\pi}{4H_{\rm O}^{2}}\ \epsilon(\hat R^{\prime}_{+}) + \right. \\
&& \,\,\,\,\, \left.- \pi \frac{(H_{\rm O}^{2}-H_{\rm I}^{2})^{2}+\kappa^{2}(H_{\rm O}^{2}+H_{\rm I}^{2})}{8\kappa H_{\rm O}^{2} H_{\rm I}^{2}} R_{0} \right] \,. 
\label{eq:wall} 
\end{eqnarray} 
The total turning point action difference between the turning point $R_0$ and $0$ is then given by adding Eq.~\eqref{eq:SBHH'} and Eq.~\eqref{eq:wall}.

The transition probability is then  determined by ${\mathcal P}\propto e^B$ with $B=2\left([I_{\rm B}+ I_{\rm W}](R_0-0)\right)$, giving

\be
\boxed{
B  =-\frac{\eta\pi}{G}\left\{ \frac{\left\{ (H_{{\rm O}}^{2}-H_{{\rm I}}^{2})^{2}+\kappa^{2}(H_{{\rm O}}^{2}+H_{{\rm I}}^{2})\right\} R_{{0}}}{4\kappa H_{{\rm O}}^{2}H_{{\rm I}}^{2}}-\frac{1}{2}\left(H_{{\rm I}}^{-2}-H_{{\rm O}}^{-2}\right)\right\} \, . } \label{eq:ItpdSdS}
\ee

Note that  $R_0$ and $I_{{\rm B}}$ and $I_{{\rm W}}$  are symmetric under the exchange $H_{\rm I}\leftrightarrow H_{\rm O}$ even though the transition rate $B$ is not symmetric. Therefore the only difference between up and down tunneling comes from the action at $\hat R=0$ which is equal to the corresponding ($\eta$ times the) entropy. Hence (choosing $\eta =+1$)

\be
\boxed{{\mathcal P}_{{\rm up}}=e^{\left(\S_{fv}-\S_{tv}\right)}{\mathcal P}_{down}} \,,\label{eq:updown}
\ee
where $\S_{fv}$ and $\S_{tv}$ refer to the entropies of the false vacuum (higher $\Lambda$) and true vacuum (smaller $\Lambda$) respectively (where the entropy for dS space is taken to be $\S=\pi/(GH^2)$ ).  So this is the statement of detailed balance and corresponds to the choice of Hartle-Hawking (HH) boundary conditions as discussed in \cite{DeAlwis:2019rxg}.
 If on the other hand we
had used the tunneling boundary conditions of Vilenkin and Linde $\eta=-1$,
we would have had to put (effectively) the coefficient $a$ in  Eq.~\eqref{eq:PsiR0} 
to zero. In this case we would have had the inverse
of detailed balance. This is
essentially because the wave function for tunneling to a
dS space from nothing is $e^{-S/2}$ rather than $e^{+S/2}$ as we
observed earlier. We speculate that this difference is due to the
fact that the HH wave function is a superposition of outgoing and incoming
waves, so is more appropriate for analysis in terms of equilibrium
thermodynamics than the tunneling wave function which is purely outgoing. For more details see~\cite{DeAlwis:2019rxg}.

\subsection*{dS to Minkowski}
For a dS to dS transition corresponding to horizons   $H_O$ and $H_I$ respectively, we can take the limit of $H_I\rightarrow 0$ to obtain the dS to Minkowski transition. Expanding in powers of 
$\varepsilon=H_I/H_O$ the turning point radius from Eq.~\eqref{eq:RoHH} is
\be
R_0\simeq \frac{2\kappa}{H_O^2+\kappa^2}\left[1+\frac{H_O^2\left(H_O^2-\kappa^2\right)}{\left(H_O^2+\kappa^2\right)^2}\varepsilon^2 \right] \,.
\ee 

Plugging this into the equation for $B$, Eq.~\eqref{eq:ItpdSdS}, and setting the limit $\varepsilon\rightarrow 0$ we get:
\be
\boxed{B=-\frac{\eta\pi}{2GH_O^2}\left[\frac{\kappa^4}{\left(H_O^2+\kappa^2\right)^2}\right]} \,,
\ee
which is the well-known dS to Minkowski result \citep{Fischler:1990pk,Coleman:1980aw,Bachlechner:2016mtp,DeAlwis:2019rxg}. Note however that this result is non-trivial since it required a cancellation between two divergent terms proportional to $1/\varepsilon^2$. In the up-tunneling case this cancellation does not occur and it can be seen as the source of the general belief that up-tunneling from Minkowski is forbidden. We will identify the source of this divergence and challenge this belief later on.


\subsection{dS to AdS}
Transitions among AdS spaces have  been studied in less detail in the Hamiltonian approach.  They can be treated in a similar way as the dS transitions changing appropriately the signs of $H^2$ in the original Eq.s~\eqref{eq:RoHH}-\eqref{eq:R'HH}. Here we will study them in more detail with the goal of writing explicit expressions for the transition rates. Let us start with dS to AdS transitions.

Let us get back to the general formulae for dS to dS tunneling, Eq.~\eqref{eq:SclassicalBulk}, that
were derived using the FMP
formalism in DMPQ (see also~\citep{Bachlechner:2016mtp}). 

On the turning point geometry $\pi_{L}=0$ the first term vanishes
and the second term is non-zero only when the argument of the inverse
cosine is $-1$. This means that the bulk integral is given by

\begin{align}
S_{{\rm B}} & =\frac{i\eta\pi}{G}\left[\int_{0}^{\hat{r}}dr\frac{dR}{dr}R\theta\left(-R'\right)+\int_{\hat{r}}^{r_{\rm max}}dr\frac{dR}{dr}R\theta\left(-R'\right)\right]\label{eq:Bulk-2}\\
 & =\frac{i\eta\pi}{2G}\left[\left(\hat{R}^{2}-R^{2}(b)\right)\theta\left(-\hat{R}'_{-}\right)+\left(R_{{\rm }}^{2}(c)-\hat{R}^{2}\right)\theta\left(-\hat{R}'_{+}\right)\right] \,.\label{eq:Bulk-3}
\end{align}
Here we have defined $r=b$ to be the lowest value of the parameter for which $R'$ is negative and $c$ is the highest value of the parameter for which $R'$ is negative. Note that the sign of $R'$ in the first line (and hence its parametrization) is determined by continuity
and the sign of $\hat{R}'_{\pm}$ which is determined by the matching
conditions and fixed by the geometry on either side of $\hat{r}$.
It should be also be noted that if the geometry on either or both
sides has horizons then $R(b)$ should be replaced by the solution
(horizon) to $A_{I}=0$ and $R(c)$ by the solution to $A_{O}=0$.

In the current case dS has a horizon $R_{{\rm D}}^{2}=H_{{\rm dS}}^{-2}$
while AdS has no horizon. Let us consider the transition $A\rightarrow B$
where $A$ is deSitter and $B$ is AdS. Thus (recall that $H^{2}\equiv\frac{8\pi G\Lambda}{3}$
is positive for dS and negative for AdS), $A_{O}=1-H_{A}^{2}R^{2}$
and $A_{I}=1-H_{B}^{2}R^{2}=1+|H_{B}^{2}|R^{2}$ and $\hat{R}_{\pm}=\frac{1}{2\kappa}\left(H_{A}^{2}+|H_{B}|^{2}\mp\kappa^{2}\right)\hat{R}$.
The last equation implies that $\hat{R}'_{-}/L>0$. Thus the first
term in Eq.~\eqref{eq:Bulk-1} is zero. Also the step function in the integral
requires $R$ to be a decreasing function of $r$ to contribute, and there is internal
horizon in empty dS (no black hole) $R(r_{\rm max})=0$.  The latter region  of course gives no contribution to the integral. Hence we have (note that from Fig.~\eqref{fig:potentialdS}, $\hat R'_+$ is negative and remains negative in the limit $M\rightarrow 0$ for all $R\geq 0$ )
\be
S_{{\rm B}}\left(\hat{R}\right)=\frac{i\eta\pi}{2G}(0-\hat{R}^{2})\,,\qquad \,S_{{\rm B}}\left(0\right)=\frac{i\eta\pi}{2G}(0) \,.
\ee

Note that a potential divergence (since AdS is non compact)  is averted since the step function in the first term of Eq.~\eqref{eq:Bulk-1} is zero.  Subtracting the second equation from the first and adding the wall contribution we thus get: 

\begin{equation}
\boxed{ 
B^{{\rm dS \rightarrow AdS}}=\frac{\eta\pi}{G}\left\{ \frac{\left\{ (H_{A}^{2}+|H_{B}^{2}|)^{2}+\kappa^{2}(H_{A}^{2}-|H_{B}^{2}|)\right\} R_{{\rm o}}}{4\kappa H_{A}^{2}|H_{B}^{2}|}-\frac{1}{2}\left(H_{A}^{-2}+|H_B^{-2}|\right)\right\} , }\label{eq:BdSAdS}
\end{equation}
with $R_{0}$ given by Eq.~\eqref{eq:RoHH} with the above substitution
$H_{B}^{2}\rightarrow-|H_{B}^{2}|$. Also in this case there is no
constraint on the tension $\kappa$. Note that this formula is in agreement with that calculated by Brown and Tetelboim using Euclidean methods. It is in fact exactly the same as the dS to dS formula with $H^2_B\rightarrow -|H^2_B|$.

As in the case of dS to dS the configuration after the transition
is actually the patching together of the original dS with an AdS
space separated by a wall. The latter will however collapse if any
matter is introduced as argued in \citep{Coleman:1980aw}. After the
collapse we will be left with a segment of dS space bounded by an
end of the world brane. On the other side of the brane there is no
geometry left and is equivalent to Witten's bubble of nothing \citep{Witten:1981gj}.

\subsection{AdS to AdS}

The AdS to AdS transitions can be analysed similarly to the previous ones.
However for transitions between AdS states there is a constraint that needs to be satisfied to guarantee that the turning point radius $R_0$ is real.
In this case $H_{I}^{2}=-|H_{I}|^{2}<0,\,H_{O}^{2}=-|H_{O}|^{2}<0$.
Now we have from Eq.~\eqref{eq:RoHH},
\[
\frac{1}{4}\left(\frac{1}{\kappa}\left(-|H_{O}^{2}|+|H_{I}^{2}|\right)-\kappa\right)^{2}>|H_{O}^{2}| \,.
\]
i.e.
\be
\kappa < \Big |\sqrt{|H_{I}^{2}|}-\sqrt{|H_{O}^{2}|}\Big | \,,\qquad {\rm or}\qquad \kappa > \Big |\sqrt{|H_{I}^{2}|}+\sqrt{|H_{O}^{2}|}\Big | \,,\label{eq:kappaconstraint}
\ee

In this case,
\be
S_B[\hat R]=\frac{\eta \pi}{2G}\left[ \left(\hat R^2-R^2(0)\right) \theta (-\hat R'_-)+\left(R^2(\infty)-\hat R^2 \right) \theta (-\hat R'_+)  \right] \,.
\ee
Also from Eq.~\eqref{eq:R'HH}  (with $H^2\rightarrow -|H^2|$) we see that for down tunneling $|H_I|^2>H_O|^2$, $\hat R'_-$ and $\hat R'_+$ are both positive for small $\kappa$, so we get $ S_B[\hat R]=S_B[0]$ and therefore the bulk contribution vanishes and the total rate comes from the wall contribution:
$I_{\rm tot}=I_{\rm W}$.

For down tunneling we then have $B=2I_{\rm W}$:

\be
\boxed{B=-\frac{\eta \pi}{2G}\left[ \frac{\left(\big |H_I^2 \big |-\big |H_O^2 \big |\right)^2-\kappa^2 \left (\big |H_I^2 \big |+\big |H_O^2 \big |\right)}{2\kappa \big |H_I^2 \big |\big |H_O^2 \big |}R_0-\left(\frac{1}{\big |H_O^2 \big |}-\frac{1}{\big |H_I^2 \big |}\right)\right]} \,.
\ee

Even though this looks very similar to the dS to dS transition case note that the sign differences and the fact that the bulk contribution vanishes make a major difference. In particular note that 
for the up-tunneling we have to change $|c_+|$ to $|c_-|$ in Eq.~\eqref{eq:wall} but also the signs of $\hat R'_\pm$ are interchanged and therefore the amplitude does not change. This means that 
\be
{\mathcal P}_{\rm up}^{{\rm AdS}\rightarrow {\rm AdS}}={\mathcal P}_{\rm down}^{{\rm AdS}\rightarrow {\rm AdS}}\,.
\ee
This is a new result and this  relation is still trivially  consistent with detailed balance if we assign zero entropy to AdS.


\subsection{Minkowski to AdS}

First let us look at the expression for the bubble radius Eq.~\eqref{eq:RoHH}. 
Taking the limit $H_{0}\rightarrow0$ we get after putting $H_{I}^{2}\rightarrow-|H_{I}^{2}|$
\[
R_{0}=\frac{2\kappa}{|-|H_{I}^{2}|+\kappa^{2}|}\left(1-\frac{H_{0}^{2}(\kappa^{2}+|H_{I}^{2}|)}{(-|H_{I}^{2}|+\kappa^{2})}+O\left(H_{0}^{4}\right)\right) \,.
\]

Now since $R_{0}\geq0$ one should take $R_{0}=2\kappa/|\kappa^{2}-|H_{I}|^{2}|+O(H_{0}^{2})$
but FMP ruled out the case $\kappa>|H_{I}|$ (in this case it turns
out that in the limit $H_{0}\rightarrow0$ the tunneling exponent
$B$ diverges), so let's focus on the case $|H_{I}|>\kappa$. Taking
the limit $H_{0}=H_A\rightarrow 0$ in \eqref{eq:BdSAdS} with
$|H_B|^2=|H_{I}|^{2}$ we get for the tunneling exponent,
\begin{equation}
\boxed{\\ \\
B=2\left( I_{{\rm tot}}|_{{\rm tp}}-\bar{I}\right)=-\frac{\eta\pi}{2G|H_{I}|^{2}}\left[\frac{2\kappa^{4}}{\left(|H_{I}|^{2}-\kappa^{2}\right)^{2}}\right] \\ } \,,\label{eq:MAdS}
\end{equation}
in agreement with~\citep{Fischler:1990pk,Coleman:1980aw}.

Let us look at this in stages separating the bulk and boundary (wall) terms. 
First we note that for $|H_{I}|^{2}>\kappa^{2}$, $\hat{R}_{\pm}^{'}>0$.
Thus we have  $I_{{\rm B}}=\frac{\eta\pi}{2G}\frac{1}{H_{0}^{2}}$
and $\bar{I}=\frac{\eta\pi}{2G}\frac{1}{H_{0}^{2}}$ so that $I_{{\rm B}}-\bar{I}=0$,
so that the bulk contribution vanishes after background subtraction
and the decay rate is fully determined by the wall contribution $ I_{{\rm W}}$ which in this limit is:
\begin{equation}
\frac{1}{2}B=I_{{\rm W}}=-\frac{\eta\pi}{4G|H_{I}|^{2}}\left[\frac{2\kappa^{4}}{\left(|H_{I}|^{2}-\kappa^{2}\right)^{2}}\right] \,,\label{eq:MAdS2}
\end{equation}
in agreement with Eq.~\eqref{eq:MAdS} as expected.

Let us note the following issue regarding these transition probabilities.  We have that
${\mathcal P}({\mathcal M}\rightarrow{\rm AdS})\sim e^{-|B|}$. This is the fact
that the transition probability goes to unity the deeper the AdS minimum
is i.e. in the limit $|H_{I}|^{2}\rightarrow\infty$. This  means the Minkowski space is unstable to decaying to the
deepest (in the EFT this would mean $|\Lambda|^{1/4}\lesssim M_{P}$
or the string or KK scale if the theory is compactified string theory).

\subsection{AdS to dS/M}
In order to avoid potential problems with the parametrization in this case, one needs to consider it as up-tunneling to dS ($A$) from an AdS black hole in the the limit $M\rightarrow 0$ ($B$). The latter is essentially the same as that studied by FMP. Since both Minkowski and AdS have no horizon the calculation in the Appendix (which is a reproduction of the FMP one) for the small mass case $iii$) applies and so from Eq.~\eqref{eq:FMP3} (setting $R_1=R_s=M=0$),

\be
\label{eq:FMP3}
S_{{\rm Bu}}\left(R_2=R_{0}\right)-S_{{\rm Bu}}(R_{1}=0)=-\frac{i\eta\pi}{G}\left(H_A^{-2}\right) \,.
\ee
In this case there is no constraint on the tension $\kappa$ and adding the wall term we get
\begin{equation}
\boxed{B^{{\rm AdS \rightarrow dS}}=\frac{\eta\pi}{G}\left\{ \frac{\left\{ (|H_{B}^{2}|+H_{A}^{2})^{2}+\kappa^{2}(-|H_{B}^{2}|+H_{A}^{2})\right\} R_{{\rm o}}}{4\kappa|H_{B}^{2}|H_{{\rm A}}^{2}}+\frac{1}{2}\left(\frac{1}{H_{A}^{2}}-\frac{1}{|H_{B}^{2}|}\right)\right\} \,, }\label{eq:BAdSdS}
\end{equation}
with $R_{0}$ again given by Eq.~\eqref{eq:RoHH} with the substitution
$H_{B}^{2}\rightarrow-|H_{B}^{2}|$.

For $|H_{B}^{2}|>|H_{A}|^{2}$ and small $\kappa$ the factor in
parentheses in the expression above for $B$ is positive, so choosing
$\eta=-1$ we get an exponentially suppressed tunneling probability
and hence an exponentially enhanced lifetime and so gravitational
collapse is exponentially more likely than tunneling to dS. However this depended on the
choice of $\eta=-1$ which is not what one chose for the
dS to dS case, where the issue was settled (as discussed in Sec.~3 of~\citep{DeAlwis:2019rxg}), by arguing that this choice (which corresponds
to the HH wave function rather than the tunneling one), gives the dominant
contribution to the wave function (and indeed was consistent with
detailed balance). Here we cannot make the same argument since that
calculation   depended crucially on the compactness of the spatial
sections of dS. 

On the other hand detailed balance holds (see below) as in the dS to dS case for the
$\eta=1$ case.  In this case this quantum transition is exponentially more probable than the gravitational collapse of AdS.
Then we have a situation where the AdS can tunnel to a configuration
of AdS separated by a wall/brane from a dS space with the AdS eventually
collapsing leaving behind a dS bounded by a end of the world brane. 

However it should be noted that the `Minkowski' limit $H_{A} \rightarrow 0$
is in fact divergent $B^{{\rm AdS \rightarrow dS}}\rightarrow\frac{\eta\pi}{2G}\frac{1}{H_{A}^{2}}\rightarrow\pm\infty$.
This is to be expected since the limit is taken from the amplitude
for transition to a dS space whose horizon and hence entropy diverges
as the dS radius goes to infinity. This is in contrast to the corresponding
up-tunneling from AdS to a Minkowski space (M) which is the limit of the AdS radius going to infinity. Note
that this limit has the same topology as M in contrast to the infinite radius limit of dS which still has the topology of a sphere.

\subsection{From nothing and back?}

In an interesting article, based on concrete cases of flux compactifications in 6D, Brown and Dahlen \citep{Brown:2011gt} have suggested interpreting
‘nothing' as the infinitely curved AdS space (to which their flux
compactified 6D theory decays to). The argument is very intuitive and explicit since they consider how the minimum of the scalar potential gets reduced by reducing the quantised fluxes and the corresponding geometry gets closer and closer to the bubble of nothing geometry of Witten  \citep{Witten:1981gj} until it reaches it in the limit of $-\infty$ cosmological constant. This interpretation is very appealing since it unifies the two concepts of nothing (the bubble of nothing and the creation from nothing). Also from the AdS/CFT interpretation, an infinite AdS curvature corresponds to a vanishing central charge which  would imply zero degrees of freedom and fits well with the concept of nothing.

This interpretation is actually
consistent with the mini-superspace ‘nothing’ which was the starting
point for the ‘no-boundary’ wave function HH or the tunneling wave
function of Vilenkin and Linde tunneling from nothing. However their
argument that up-tunneling from AdS to dS/M is prohibited (based on
the non-compactness of the spatial sections of AdS)\footnote{In any case the argument depended on not including the Gibbons-Hawking regulator term as in the Euclidean arguments mentioned earlier.} is not necessarily valid since
as we saw earlier the FMP bulk contribution is zero at the turning
points so that the tunneling amplitude is actually finite. To see
this let us take the limit $|H_{B}^{2}|\rightarrow\infty$ first in Eq.~\eqref{eq:RoHH} (with $H_{O}^{2}\rightarrow-|H_{B}^{2}|$ which
gives $R_{0}\rightarrow2\kappa/|H_{B}^{2}|$) and then substituting
in Eq.~\eqref{eq:BAdSdS} we get
\[
B^{{\rm AdS \rightarrow dS}}\rightarrow\frac{\eta\pi}{G}\left\{ \frac{\left\{ (|H_{B}^{2}|)^{2}\right\} 2\kappa/|H_{B}^{2}|}{4\kappa|H_{{\rm B}}^{2}|H_{A}^{2}}+\frac{1}{2}\left(\frac{1}{H_{A}^{2}}+0\right)\right\} =\frac{\eta\pi}{2G}\frac{1}{H_{A}^{2}} \,.
\]
 That is if we define as nothing the limit of AdS with $|H_B|\rightarrow \infty$. We get:
 \be
 \boxed{B^{{\rm Nothing \rightarrow dS}}=\frac{\eta\pi}{2G}\frac{1}{H_{A}^{2}}} \,.
 \label{eq:nothingrate}
 \ee
This is precisely the (log of the) Hartle-Hawking (for $\eta=+1$) or the Vilenkin-Linde (for $\eta=-1$) tunneling factor for creating a universe from nothing.

Thus, we agree with the proposal of \citep{Brown:2011gt} to identify the two definitions of nothing, the limit of infinite curvature AdS as representing the bubble of nothing and the nothing of Vilenkin or Hartle-Hawking regarding the wave function of the universe interpretation as creation from nothing. But contrary to the claim of \citep{Brown:2011gt} in which creation from nothing does not happen,  we can reproduce
the tunneling from nothing picture  by interpreting nothing
as deep AdS as they did. It is interesting to note that even though
the bubble radius goes to zero in this limit (which normally would
have been interpreted as signalling the absence of tunneling) the singularity in $B/2$ cancels resulting in a finite tunneling probability.

We may question the validity of taking the limit $|H_B|\rightarrow \infty $ since the EFT is only valid up to energies smaller than the Planck mass. But we can reproduce this result as the leading term in an expansion in powers of $\varepsilon^2=H_A^2/|H_B|^2$ and $\delta^2=\kappa^2/|H_B|^2$ with $\varepsilon,\delta \ll 1$ but still keeping $|H_B|\leq M_P$.

\subsubsection*{Detailed balance in dS/AdS transitons.}

The results of the above subsections shows that detailed balance holds
for dS to and from AdS transitions provided we assign zero entropy to empty Anti-deSitter space (as one should expect
given that empty AdS has no horizon).  In this case we have,
\begin{equation}
\frac{P^{{\rm AdS \rightarrow dS}}}{P^{{\rm dS \rightarrow AdS}}}=\frac{e^{B^{{\rm AdS \rightarrow dS}}}}{e^{B^{{\rm dS \rightarrow AdS}}}}=\frac{\exp\left(\frac{\eta\pi}{2G}\frac{1}{H_{A}^{2}}\right)}{\exp\left(-\frac{\eta\pi}{2G}\frac{1}{H_{A}^{2}}\right)}=e^{\eta\left(S_{{\rm dS}}-\left(S_{{\rm AdS}}=0\right)\right)} \,,\label{eq:dSAdSdb}
\end{equation}
which is the statement of detailed balance (taking $\eta=+1$).  Note that the
result holds also in the limit $H_{B} \rightarrow 0$ which is the Minkowski
limit of AdS so that detailed balance holds for transitions between
dS and Minkowski space-times provided of course that we assign zero
entropy to empty Minkowski space. 

\subsubsection*{Comparison to Euclidean methods}

In the Euclidean calculations (CdL, BT) it appears that up-tunneling
from AdS (or M) to dS is forbidden. However that is because in those
calculations the background action is taken to be proportional to
the volume of AdS space - which is infinite. However as we see from Eq.~\eqref{eq:BAdSdS} in the WKB calculation the initial point is
$\hat{R}=0$ so we had $S_{{\rm B}}\left(0\right)=0$ rather than
infinity. This is also consistent with taking the limit of the black hole mass to zero in the SAdS to dS calculation of Sec.~\ref{sec:sdstosds}.

The difference in fact corresponds to the different assignments of
entropy to AdS space. Our calculations (both the direct AdS to dS
and the SAdS to dS) effectively assigned zero entropy to empty AdS
space. The Euclidean calculation on the other hand effectively assigned
infinite entropy to empty AdS.


\section{Transitions from Black Hole Backgrounds}

Let us now consider the most general situation allowed by spherical symmetry, namely that the solution to Einstein's equations include a mass parameter $M$ that appears as an integration constant in the solution of the Hamiltonian constraints that corresponds to a black hole mass. Since the Schwarzschild solution implies the existence of a horizon considering $M\neq 0$ makes an important difference.\footnote{Note that in following the Hamiltonian approach,  which at this point is amenable to computation only in the extreme thin wall (brane) approximation, we have nothing to add to the current debate regarding the impact of black holes in Higgs decay~\cite{Gregory:2013hja,Burda:2015isa, Canko:2017ebb, Strumia:2022jil}.}

\subsection{SdS to SdS}

As discussed above,the classical turning points for the geometry occur at $\pi_{L}=0$,
i.e. 
\[
\frac{R'^{2}}{L^{2}}=A(R)=1-\frac{2MG}{R}-H^{2}R^{2} \,.
\]
When $r=\hat{r}$ these are the turning points for the brane i.e.
the solutions of $V=-1$. For $3\sqrt{3}GM<H^{-1}$ the geometry has
two horizons $R_{b}<R_{c}$. We may identify $R_{s}$ as the black hole
(Schwarzchild) horizon (becoming $2GM$ when $H\rightarrow0)$ while
$R_{c}$ is the cosmological horizon (becoming $H^{-1}$ when $M\rightarrow0$).
We have
\begin{align*}
A =-\frac{H}{R}\left(R-R_{-}\right)\left(R-R_{b}\right)\left(R-R_{c}\right) \,,\\
R_{-}<0 <2GM<R_{b}<3GM<R_{c} \,.
\end{align*}
We also see from the turning point equation $V=-1$ that
\[
R_{b}<R_{1}<R_{2}<R_{c} \,.
\]
For the turning point geometries the bulk action $S_{B}$ simplifies
with the first term in Eq.~\eqref{eq:SclassicalBulk} giving zero and the second term
contributes only when $\epsilon(R')=-1$ i.e. whenever $\cos^{-1}\left(\frac{R'}{L\sqrt{A_{{\rm I,{\rm O}}}}}\right)=\pi$.
Thus we have
\begin{eqnarray}
iS_{{\rm B}}(\hat{R}) & =- & \frac{\eta\pi}{G}\left[\int_{0}^{\hat{r}}drR'R\theta(-R_{-}')+\int_{\hat{r}}^{r_{\rm max}}drR'R\theta(-R_{+}')\right]\nonumber \\
 & = & \frac{\eta\pi}{G}\left[\int_{0}^{\hat{R}}dRR\theta(-R_{-}')+\int_{\hat{R}}^{R(r_{\rm max})}dRR\theta(-R_{+}')\right] \,.\label{eq:SBtp}
\end{eqnarray}
The general dS-black hole case is complicated. It is discussed in detail in the appendix - see Sec.~\eqref{sec:sdstosds}.

For the small black hole case the direct calculation in a static patch between the two horizons gives \begin{align}
I_{{\rm {\rm Bu}}}\left[\hat{R}\right]  =\frac{\eta\pi}{2G}\left[(\theta(-\hat{R}'_{+})-\theta(-\hat{R}'_{-})\hat{R}^{2}+\theta(-\hat{R}'_{-})\left(R_{I,c}^{2}\right)-\theta(-\hat{R}'_{+})\left(R_{O,b}^{2}\right)\right]
  \,.\label{eq:Ibulk0}
\end{align}

Hence we get after adding the wall contribution
(see Eq.~\eqref{eq:IAB})
\begin{equation}
I^{AB}\left[R_{2}-R_{1}\right]=\frac{\eta\pi}{2G}\left[\left(R_{{\rm c}}^{B}\right)^{2}-\left(R_{{\rm {\rm b}}}^{A}\right)^{2}\right]+I_{W}^{AB}\left[R_{2}-R_{1}\right] \,.\label{eq:IAB}
\end{equation}

Let us now consider the transitions $A\rightarrow B$ (with $S_{A}>S_{B}$)
which we may call up-tunneling and $B\rightarrow A)$ we call down-tunneling. Noting that the wall action $I_{W}$ is symmetric under the interchange of  $A$ and  $B$.\footnote{While the bulk action is symmetric only for a general point and loses this symmetry at the turning points $R_{1,2}$ the wall action clearly has this symmetry even at these points as
is seen from Eq.~\eqref{eq:WIO}.} Hence  we get, 
\begin{equation}
\frac{P_{\uparrow}}{P_{\downarrow}}=e^{\frac{\pi}{G}\left[\left(R_{B,c}^{2}-R_{A,b}^{2}\right)-\left(R_{A,c}^{2}-R_{B,b}^{2}\right)\right]}=e^{\S_{B}-\S_{A}}.\label{eq:detalied balcance}
\end{equation}

Defining the total entropy of SdS space as $S\equiv \frac{\pi}{ G}(R^2_c+R^2_b)$ (i.e. the sum of the  cosmological and black hole horizon entropies).
Thus we have non-trivially obtained detailed balance again.

In the limit $R_{A,Bs}\rightarrow0$ we recover the earlier results
for $dS$ to $dS$ transitions. One may think that to get FMP/FGG
we need to take $R_{A,c}\rightarrow\infty$. This would give $P_{\uparrow}\rightarrow0$
in agreement with the corresponding limit in the dS to dS case. The
FMP/FGG case however corresponds to subtracting the infinity in the
horizon area term of dS/SdS when the dS radius goes to infinity to
get the entropy of the black hole in asymptotically Minkowski space
to be just the black hole entropy. 

The difference comes from the fact that in the asymptotically Minkowski
case Gibbons and Hawking \citep{Gibbons:1976ue} added a term (to
the boundary term which is necessitated by Dirichlet boudary conditions)
evaluated in flat space which cancels the otherwise infinite contribution
of the GHY term in the asymptotic limit. This infra-red subtraction
is what gives the entropy of the black hole to be $\pi r_{s}^{2}$
and the entropy (as well as the ADM energy) of flat Minkowski space
to be zero.\footnote{In \cite{Weinberg:2012pjx} page 283 it is stated the GHY term for Minkowski is negative infinity and is zero in the absence of the GHY term. However it should be noted that the GHY term consists of two pieces and the second piece is explicitly included to make the Minkowski vacuum action (and hence its entropy) zero. A similar term should be included if the AdS vacuum is to be assigned zero entropy since it has no horizons. This means that the entropy of both a Minkowski black hole and an AdS black hole is just the horizon entropy of the black holes.} This is a physical requirement. In other words the infinite
radius limit of dS space is not flat Minkowski space any more than
a topological 3-sphere of arbitrarily large radius is the same as
$R^{3}.$ The topology of dS is $R\times S_{3}$ and however large
its radius, is not $R^{4}$. Thus  the above formulae - although
valid for SdS spaces of arbitrary radii, do not imply (from the vanishing
of Eq.~\eqref{eq:detalied balcance}) that they forbid up-tunneling from
asymptotically Minkowski space. 

In fact it has been argued by many authors (perhaps the earliest were
\cite{Banks:2000fe,fischler2000taking,Fischler:1989se}), that the horizon
entropy of dS space is the maximum entropy that this space can hold.
With this interpretation then, in the limit of the horizon radius
going infinity, the ensuing infinite entropy of Minkowski space, should
also be interpreted as the maximum entropy that this space can contain,
which of course is reasonable. It is not therefore the entropy of
the Minkowski vacuum, which should indeed be zero.

As we will discuss soon, the
same is true of SAdS to SdS (or dS) since the entropy of SAdS is simply
the entropy of the black hole since AdS has no cosmological horizon.

However for large and intermediate black hole masses (as defined in Sec.~\ref{sec:sdstosds} we do not have detailed balance. This is also the situation for the FMP case reviewed in the appendix Sec.~\ref{sec:stods}.

\subsection{Schwarzschild (A)dS to Schwarzschild (A)dS transitions}

Let us now consider the generic case in which the mass parameters $M_{\rm O}$ and $M_{\rm I}$ are non-vanishing. We will then reconsider the forbidden up-lifting configurations and illustrate that in the case $M_{\rm O}\neq 0$ some forbidden transitions become possible. The main reason for this change is the existence of a horizon and the fact that there are two rather than one turning points.

\subsubsection{Black hole to dS - the FGG transition.}

We will start reviewing the original up-tunneling proposal of Farhi, Guth and Guven (FGG) \cite{Farhi:1989yr} and reconsidered in \cite{Fischler:1990pk} in which the background is asymptotically flat Schwarzschild black hole.

The brane tunnels through the
potential and the probability of transition
from the black hole state to the one with a dS space 
is given essentially by the relative probability for being at the
turning point $R_2$ to that of being at the turning point $R_{1}$,
i.e.
\begin{equation}
{\mathcal P}(BH\rightarrow{\cal N)}=\frac{|\Psi(R_2)|^{2}}{|\Psi(R_{1})|^{2}},\label{eq:Pio}
\end{equation}
with $R_{2,1}$ being the right and left turning points.

In the FGG/FMP process we have a transition from a Schwarzschild
black hole to a de Sitter space (adjoined to the hole with a brane
at the junction). The integrals in Eq.~\eqref{eq:Integrals} for $H_{\rm O} = 0$ but $M_{\rm O}\neq 0$ cannot be done analytically but we can approximate them for a small black hole mass. The important point is that  now the background contribution is determined not by the entropy of dS but by the black hole entropy:

\be
\bar I=S_{BH}=\frac{\pi R_S^2}{G}=4\pi M_{\rm O}^2 \,.
\ee

Therefore for a small mass black hole $M_{\rm O}<M_{D}\equiv\frac{H^{2}-\kappa^{2}}{2GH^{3}}\simeq\frac{M_{P}^{2}}{H}$
we can neglect the $M_{\rm O}$ dependence in the integrals in Eq.~\eqref{eq:Integrals} and with the background $\bar I$ given as above we have,
\be
{\mathcal P}(BH\rightarrow dS)\simeq e^{\frac{\pi}{G}(R_{dS}^{2}-R_{{\rm Schwazchild}}^{2})+2I_{{\rm brane}}}\simeq e^{S_{dS}-S_{BH}} \simeq e^{S_{dS}}\neq 0 \,.\label{eq:bhds}
\ee

The important point is that, contrary to the limit $H_{\rm O} = 0$ for up-tunneling dS to dS which has a vanishing transition rate, now the transition rate from Schwarzschild black hole to dS is non-vanishing even for small values of $M_{\rm O}$.

Note  here the entropy of the final state is much higher than the
black hole entropy, so this is consistent with increasing entropy.

Let us contrast the situation here with that obtained earlier of dS to dS transitions.
In~\cite{DeAlwis:2019rxg} we have argued that the FGG/FMP starting configuration is a state with a definite energy (that of the black hole) and that the thermodynamic entropy of the initial state should be identified with the log of the degeneracy of the Hilbert space of the black hole and the background flat space being treated as the Minkowski vacuum identified as a unique state. in contrast to this 
in the dS case it has been argued that its entropy is maximal and indeed is supposed to correspond to the dimension of the Hilbert space that can be accommodated on the horizon. The latter obviously goes to infinity at infinite radius but this should not be confused with the entropy of empty Minkowski space.

\subsection{Schwarzschild AdS to dS}

For Schwarzschild AdS to dS we follow the same steps as FMP did for Schwarzschild to dS. The two turning points $R_{1}, R_{2}$ for which $V=-1$ are in between the dS radius $R_{\rm dS}=1/H_I$ and the Schwarzschild radius $R_{\rm S}$ that is determined by solving the cubic equation:
\be
H_{\rm O}^2R_{\mathcal S}^3+R_{\mathcal S}-2GM=0 \,,
\ee
namely
\be
R_{\mathcal S}\leq R_{1} < R_{2} \leq R_{\rm dS} \,.
\ee
This can be easily seen by the conditions coming from the expression of the potential at the turning points $\hat A_I=1-H_I^2R^2>0$ implying $R_{2,1}\leq R_{\rm dS}$ and $\hat A_O=1-2GM/R+H_{\rm O}^2R^2 >0$ which by looking at the coefficients determining the single root of  the cubic, it can seen that it implies $R_{\mathcal S}\leq R_{{1, 2}}$.

Also to define the domains for the integration parameter $M$ we compute it first for the case $R'_+=0, V=-1$ which implies gives the value for $M$:
\be
M=M_{\rm S} =\frac{H_{\rm O}^2+H_{\rm I}^2+\kappa^2}{2G\left(H_{\rm I}^2+\kappa^2 \right)^{3/2}}, \qquad {\rm for} \qquad R'_+=0, V=-1 \,.
\ee
For the case $R'_-=0, V=-1$ we get
\be
M=M_{\rm D}=\frac{H_{\rm O}^2+H_{\rm I}^2-\kappa^2}{2GH_{\rm I}^3},  \qquad {\rm for} \qquad R'_-=0, V=-1 \,.
\ee
Note that for $H_{\rm O}=0$ this reduces to the FMP results as it should. It is also easy to prove that, as in the FMP case, $M_{\rm D}\leq M_{\rm S}$.

Therefore we have the same situation as in the Schwarzschild to dS transition in  which the bulk contribution to the transition rate is determined by:

\bea
I_{{\rm B}}\bigg|_{\rm tp} \equiv I_{{\rm B}}\bigg|_{R_{{\rm 1}}}^{R_{\rm 2}} =\begin{cases}
\frac{\eta\pi}{2G}(R_{2}^{2}-R_{{1}}^{2})\,, \qquad & M > M_{\rm S} \,,\label{eq:SB1}\\
\frac{\eta\pi}{2G}(R_{2}^{2}-R_{\mathcal S}^{2})\,, \qquad & M_{\rm S} > M > M_{\rm D} \,,\label{eq:SB2}\\
\frac{\eta\pi}{2G}(R_{\rm dS}^{2}-R_{\mathcal S}^{2})\,, \qquad & M_{\rm D} > M \,.\label{eq:SB3}
 \end{cases}
\eea

The relevant figure for this is the same as in the FMP case i.e. Eq.~\eqref{fig:potentialdS} in the appendix, since both Minkowski and AdS have no cosmological horizons.

As in FMP we are interested in the latest case $M_{\rm D} > M$ to take the small $M$ limit. As in there we may take the limit $M\rightarrow 0$ \footnote{This limit is well defined as long as we stay away from the horizon. The squared curvature tensor is $\propto (GM)^2/R^6$ which goes to zero as $M\rightarrow 0$ unless $R=2GM$.}.

So we have explicitly a non zero transition rate from AdS black hole to dS which is interesting by itself. The interesting questions to ask is how the whole transition rate depends on the values of the parameters $M,H_{\rm I}, H_{\rm O}, \kappa$. In particular if it prefers transitions to smaller or higher values of $H_{\rm I}$ for a fixed $H_{\rm O}$ or viceversa. Also to analyse the transition rate in the extreme cases $H_{\m I,O}\rightarrow M_p$ from below and $M\rightarrow M_p$ from above. This should be done numerically combining the bulk and the wall contributions to the transition rate.

\subsection{Vacuum transitions in quantum gravity and thermodynamics\label{sec:susskind}}

In this subsection we address the interesting points made in \citep{Susskind:2021yvs} regarding the viability of the FGG/FMP process.  Susskind argues that the FGG/FMP process should be forbidden since it violates
what he calls the central dogma  of gravitational systems with horizons
- specifically to de Sitter-like geometries.

\textit{As seen from a causal patch a cosmological space-time can
be described in terms of an isolated quantum system with $\text{{\rm Area}}/4G$
degrees of freedom, which evolves unitarily under time evolution.}

In the next paragraph of that paper the author  assumes that an observer in a causal patch sees a world of finite
entropy satisfying the second law of thermodynamics.

Now, while unitarity of the S-matrix can be shown to lead to the second
law for non-gravitational systems (for a proof see for instance \citep{Weinberg:1995mt}
chapter 3) it is far from clear (and interesting to explore) how to extend this to gravitational
systems particularly transitions where space-time itself changes.
Furthermore there is the notorious problem of time in quantum gravity.
So even though the first statement above holds referring to black holes
in asymptotically flat spacetime (where the notion of a S-matrix may be
formulated) it is far from clear how these statements apply for say
dS space or SdS space. 

Nevertheless Susskind proceeds to conjecture
that they do. Indeed as he says ``The black hole version of this
dogma is, with good reasons, widely accepted. Less is known about
cosmological horizons, so we should consider the cosmological version
to be a conjecture.'' Furthermore referring to the FGG process ``Because
this example involves following a causal patch as it falls behind
a black hole horizon, there may be reasons to be less certain about
the application of the cosmological dogma, but I will assume it is
valid.''

Unlike FGG/FMP who discuss an eternal black hole space-time which
nucleates a spherical wall/brane whose inside is a portion of a dS
space, Susskind's version as presented in section 4, assumes a background
dS space in which a black hole (of mass $M$) is nucleated as a thermal
fluctuation with (Boltzmann) probability $P_{bh}\propto e^{-M/T_{A}}=e^{-2\pi MR_{A}}.$
Here $T_{A}=1/2\pi R_{A}$ is the horizon temperature of the background
dS space. So Susskind's interpretation of FGG is that they predict
an up-tunneling rate for the transition $A\rightarrow B$ (where $B$
is a dS space with smaller horizon radius ($R_{B}<R_{A})$ and hence
smaller entropy) 

\begin{equation}
\bar{\Gamma}_{up}=e^{-2\pi MR_{A}}\Gamma_{FGG} \,.\label{eq:Gammabar}
\end{equation}
On the other hand (according to Susskind) detailed balance requires
the up-tunneling rate to be 
\begin{equation}
\Gamma_{up}=e^{\S_{B}-\S_{A}}\Gamma_{CDL} \,,\label{eq:upCDL}
\end{equation}
where $\Gamma_{CDL}$ is the Coleman-DeLuccia down-tunneling rate
and $\S_{A}=\pi R_{A}^{2}/G,\,\S_{B}=\pi R_{B}^{2}/G$. Since $M,\S_{B},\Gamma_{FGG}$
and $\Gamma_{CDL}$ are all independent of $R_{A}$ for $R_{A}\rightarrow\infty$
it was argued that in this regime
\begin{equation}
\frac{\bar{\Gamma}_{up}}{\Gamma_{up}}\approx\frac{\Gamma_{FGG}}{\Gamma_{CDL}e^{\S_{B}}}e^{\S_{A}}\sim e^{\S_{A}} \,,\label{eq:Susskind}
\end{equation}
in violent disagreement with detailed balance.

Several comments are in order:
\begin{itemize}
\item The Hamiltonian calculation of dS$_{A}\rightarrow$ dS$_{B}$ using the
FMP\citep{Fischler:1990pk} formulation gives exactly Eq.~\eqref{eq:upCDL}
\citep{DeAlwis:2019rxg} so this quantum mechanical calculation is
not in conflict with detailed balance. In this sense we address the main criticism of Susskind for the process to occur.

\item In the presence of a black hole the issue is more complicated. The
original calculation of FGG and FMP discussed the nucleation of a
brane with a dS space inside it and an asymptotically flat Schwarzschild space outside. Here there are three cases~\citep{Fischler:1990pk}\footnote{Actually there is a discrepancy between FMP and FGG here. The relevant equation in the latter \citep{Farhi:1989yr} is Eq.~(5.34) which is not quite the same as the equations 48 plus 49 of \citep{Fischler:1990pk}. In particular for $M>M_{S}$ FGG gives $I=I_{W}$ whereas FMP give
$I=I_{W}+\frac{\pi}{2G}\left(R_{2}^{2}-R_{1}^{2}\right)$. See appendix
for notation.} as discussed in detail in the appendix  Sec.~\ref{sec:stods}. The first two cases (see Eq.s~\eqref{eq:FMP1}, \eqref{eq:FMP2}) clearly do not satisfy detailed balance, but the third case $M<M_{D}<M_{S}$ gives  Eq.~\eqref{eq:upCDL} \textit{provided} we identify $\S_{A}=\pi R_{S}^{2}/G$ i.e the entropy of a black hole in flat space as the entropy of the initial state. All this assuming also that down tunneling is given by CDL which is only true in the limit where the black hole mass tends to zero.

\item Actually the relevant comparison to the FMP type calculation that should be made is to an SdS to SdS transition as discussed in detail in the Appendix. Here there are also  three cases with the case of small black holes giving Eq.~\eqref{eq:upCDL} with $\S_{A,B}=\frac{\pi}{G}\left(R_{A,B}^{2}+\left(R_{A,B}^{{\rm BH}}\right)^{2}\right)$ with the down tunneling rate modified from CDL to incorporate the black holes. In the limit where the black hole masses go to zero this
is exactly Eq.~\eqref{eq:upCDL}.

\item On the other hand as in the asymptotically flat case there are two other cases (see appendix) corresponding to different ranges of black hole masses,    which do not seem to satisfy detailed balance. The interpretation of these regimes is not clear to us.
\end{itemize}
To summarize: the Hamiltonian calculation shows that  transitions
between space-times with horizons such as dS to dS, and SdS to SdS
with small mass black holes, do satisfy detailed balance and may not be in conflict with Susskind's reasoning. However there
are also cases with large or intermediate mass black holes (as defined in the appendix), that do not satisfy Susskind's central dogma. Since the dogma in the case of tunneling of space-times through barriers, as he himself says, may not apply, this should not be surprising. A proper understanding of the difference of the mass ranges regarding detailed balance would be interesting.

A last comment; from holography it is expected that the empty AdS has zero entropy.  To see this let us consider the eternal AdS black hole~\cite{Maldacena:2001kr}.   Empty AdS can be obtained by taking the mass of the black hole to zero.  Doing this so also the radius, the area and the entropy of the black hole vanish.  If there is a CFT, as in the eternal AdS case, measuring the number of degrees of freedom, of the system,  then these will also  decrease up to the point when there is no black hole and only one state is left which means vanishing entropy.  It will be interesting to further explore this issue.


\subsection{Brane trajectory after nucleation}
In this section we will study the trajectory of the brane after the nucleation.  We will focus on transitions where the true vacuum is dS.  Following~\cite{Blau:1986cw} the geometry of the brane can be studied by solving the equations for the brane position $\hat{R}$.  For the readers' convenience we repeat the formulae for the brane motion.  The metric on the brane is 
\begin{equation}
ds^{2}=-dt^{2}+\hat{R}^{2}(t)d\Omega^{2} \,.\label{eq:brane-metric}
\end{equation}
the first (energy) integral of the equation of motion for the brane is (from now on we will drop the hat on $R$ in this section
since we are just discussing the brane motion),
\begin{equation}
\dot{R}^{2}+V=-1 \,,\label{eq:brane-eom}
\end{equation}
 where the potential may be written as
\begin{equation}
V=-\frac{1}{\left(2\kappa R\right)^{2}}\left(\left(A_{I}-A_{O}\right)+\kappa^{2}R^{2}\right)+A_{I}-1.\label{eq:V1}
\end{equation}
In the case of interest (i.e. AdS black hole to dS), 
\[
A_{I}=1-H_{I}^{2}R^{2},\,\,A_{O}=1-\frac{2GM}{R}+H_{O}^{2}R^{2},
\]
where $H_{I}$ is the inverse of the de Sitter radius and $H_{O}$
is the inverse of the AdS radius.  Substituting these into Eq.~\eqref{eq:V1} we get
\begin{equation}
V=-\frac{R^{2}}{R^{*2}}+\frac{\alpha}{R}-\frac{\beta}{R^{4}} \,,\label{eq:V2}
\end{equation}
where 
\begin{equation}
\alpha=4\left(H_{O}^{2}+H_{I}^{2}-\kappa^{2}\right)\frac{GM}{\left(2\kappa\right)^{2}}\,, \qquad \beta=\frac{\left(2GM\right)^{2}}{\left(2\kappa\right)^{2}}>0 \,,\label{eq:alpha-beta}
\end{equation}
and
\begin{equation}
\frac{1}{R^{*2}}=\frac{\left(H_{O}^{2}+H_{I}^{2}\right)^{2}+2\kappa^{2}\left(H_{I}^{2}-H_{O}^{2}\right)+\kappa^{4}}{\left(2\kappa\right)^{2}}\,.\label{eq:R*}
\end{equation}
Note that the equation $dV/dR=0$ has only one real root and that
$V\rightarrow-\infty$ for both $R\rightarrow0$ and $R\rightarrow\infty$,
so the potential rises from negative infinity to a maximum below zero
(recall that $V<0$ for all $R$) and then falls back to negative
infinity just as in the BGG case~\citep{Blau:1986cw}. Depending on the value of $M$ we have the three cases discussed earlier.

To compute the motion of the brane in (say) the dS metric inside the
bubble we choose (conformal) global coordinates
\begin{align}
ds^{2} & =\frac{1}{H_{I}^{2}\cos^{2}T}\left(-dT^{2}+dr^{2}+\sin^{2}rd\Omega^{2}\right) \,,\label{eq:global-dS}\\
T_{0} & \le T<\frac{\pi}{2},\,r_{0}\le r<\frac{\pi}{2}\,,\nonumber 
\end{align}
where $T_{0},r_{0}$ are the coordinte time and radius at which the
bubble is nucleated. Embedding the brane metric Eq.~\eqref{eq:brane-metric} in the above we get (denoting $\dot{X}\equiv\frac{dX}{dt}\,X'=\frac{dX}{dr}$),
\begin{align}
T & =T(t),\,r=r(t),\,R=\frac{\sin^{2}r}{H_{I}^{2}\cos^{2}T} \,,\label{eq:embedding1}\\
\dot{T} & =\frac{H_{I}\cos T}{\sqrt{1-r'^{2}}},\,\dot{r}=\frac{H_{I}\cos T}{\sqrt{1-r'^{2}}}r' \,.\label{eq:embedding2}
\end{align}
We also have
\begin{align*}
\dot{R} & =\frac{\partial R}{dT}\dot{T}+\frac{\partial R}{dr}\dot{r}=\frac{\dot{T}}{H_{I}}\left(\tan T\sec T\sin r+\frac{dr}{dT}\sec T\cos r\right)\\
 & =\frac{1}{\sqrt{1-r'^{2}}}\left(\tan T\sin r+r'\cos r\right) \,.
\end{align*}
Hence using Eq.s~\eqref{eq:brane-eom}, \eqref{eq:V2} and \eqref{eq:embedding1}
we have 
\begin{equation}
\frac{1}{\sqrt{1-r'^{2}}}\left(\sin T\sin r+r'\cos r\cos T\right)=\sqrt{\frac{\sin^{2}r}{H_{I}^{2}R^{*2}}-\alpha H_{I}\frac{\cos^{3}T}{\sin r}+\beta H_{I}^{4}\frac{\cos^{6}T}{\sin^{4}r}-\cos^{2}T} \,.\label{eq:brane-eom2}
\end{equation}
The small black hole regime may be defined as $\alpha H_{I}\ll1,\,\beta H_{I}^{4}\ll1.$
In the limit $M\rightarrow0$ this obviously tends to the formal AdS
to dS result.

Note that as $T\rightarrow\pi/2$ (i.e. in the infinite future) we
get $\sqrt{1-r'^{2}}=H_{I}R^{*}\ne0$. i.e. the brane speed does not
reach the speed of light, as in the BGG case.

As is now well known \cite{Cespedes:2020xpn},  in the case of dS/dS transition,  the wall after nucleation follows a geodesic in $SO(3,1)$.  This has been to argued that the the spacetime after the foliation is open since it fits naturally with the picture of the wall never crossing the horizon.  The argument for the wall trajectory relied heavily in the fact that the tangent acceleration  in the dS/dS  case is always constant.  Here we show that adding a mass to the exterior vacuum changes this pictures allowing for different trajectories of the wall.  First,  let us note that the acceleration of the wall can be computed from the junction conditions to be~\cite{Blau:1986cw},
\begin{align}
K_{\tau\tau}=-\frac{\ddot R-H^2 R}{\sqrt{1-H^2R^2+\dot R^2}} \,,
\end{align}
whose value can be inferred by solving the potential equation for the junction conditions in Eq.~\eqref{eq:EofM}.
In the case of dS/dS,  the solution can be obtained analytically,  $R=R_0\cosh(t/R_0)$ and the acceleration is constant.  In  general when there is a mass this does not hold,  and we find that the acceleration is given by
\begin{align}
-K_{\tau\tau}=\frac{\frac{(1-H_0^2 R_*^2)}{R_0^2}-\frac{\alpha }{2R(\tau)^3}-\frac{2\beta }{R(\tau)^6}}{\sqrt{\frac{(1-H_0^2 R_*^2)}{R_0^2}-\frac{\alpha }{R(\tau)^3}-\frac{\beta }{R(\tau)^6}}} \,,
\end{align}
where we have used the equation of motion. $\dot R^2+V=-1$,  Expanding for small $\alpha H_I$ and $\beta H_I^4$,  we can write the LHS as,
\begin{align}
\frac{\ddot R-H^2 R}{\sqrt{1-H^2R^2+\dot R^2}}=\frac{\sqrt{1-H_0^2 R_*^2}}{R_*}-\frac{R_*}{\sqrt{1-H_0 R_*^2}}\left(\frac{\alpha}{R^3}-\frac{5}{2}\frac{\beta}{R^6}+\mathcal{O}(R_0^3\alpha^2/R^6)\right) \,.
\end{align}
From this we see that for small radius the wall initially  accelerates/decelerates until $R(\tau)$ grows large enough such that the acceleration asymptotes to a constant.  Notice that in the case that the mass in the outer region is zero both $\alpha$ and $\beta$ are zero in which case the acceleration is always constant as we anticipated.  Another feature is that this results do not depend on the outer region being SdS or AdS and hence this is a feature of the outside region having a mass. 

\begin{figure}[h!] 
\begin{center} 
\includegraphics[scale=0.55,trim=2.1cm 8cm 1cm 5cm,clip]{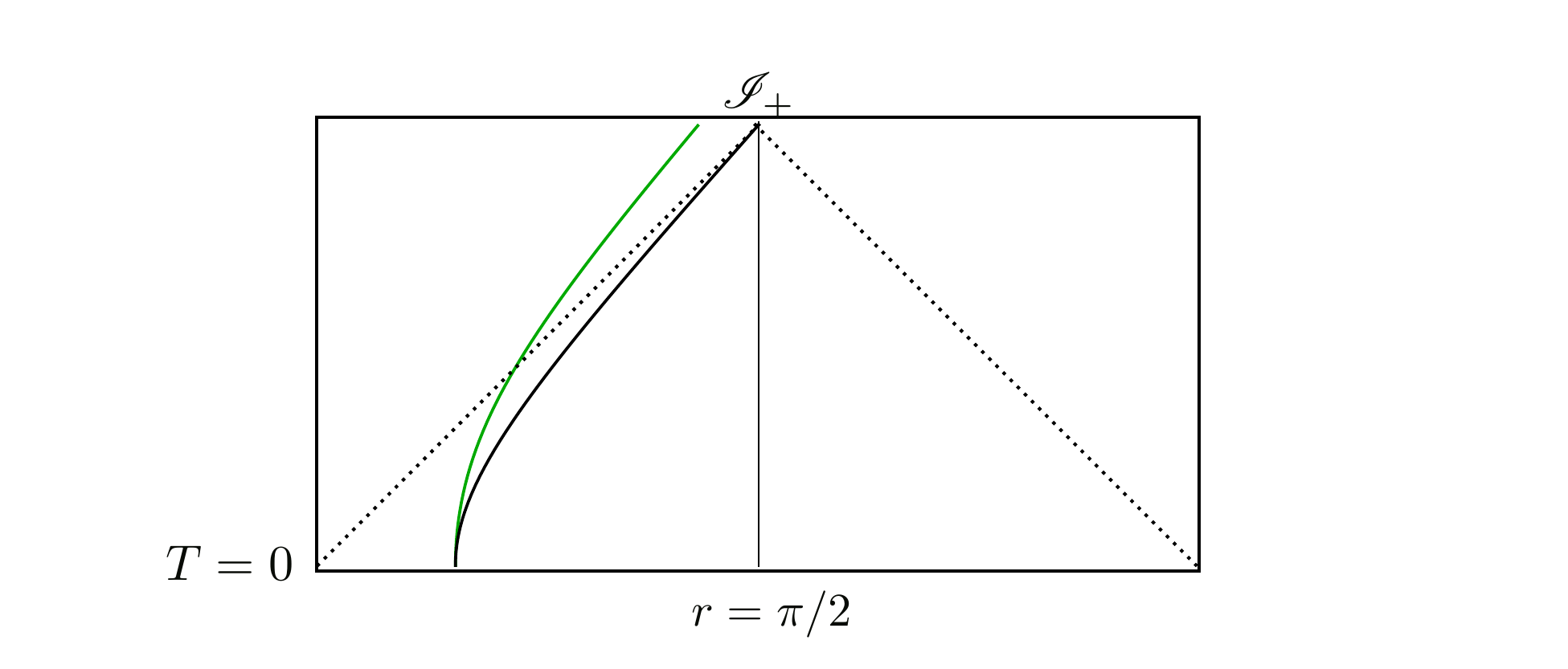}
\caption{\footnotesize{Penrose diagram of the wall trajectory. The green line is the solution in the case of $\alpha=.11H$, $\beta= 0.0014 H^4$, $R_*=0.55 H$.  The black line corresponds to the case whe $M=0$ annd $R_*=0.55 H$.  Note that the green line crosses the dotted line meaning that the wall crosses the horizon of the De Sitter observer } \label{fig:WallTrajectory}}
\end{center} 
\end{figure}
One of the consequences of the change of acceleration is that the wall can now enter into the comoving horizon of an observer on one of the hemispheres.  

This was shown by writing the radius of the wall in global coordinates $R(T)=a(T)\sin(r(T)$ as above.
 We plot the solutions to Eq.~\eqref{eq:brane-eom2} where we can see that for certain values of $\alpha$ the wall indeed crosses the lightcone $r=T$ as can be seen from Fig.~\ref{fig:WallTrajectory}.  Clearly this is due to the fact that the worldsheet of the trajectory  only has $O(3)$ symmetry and hence it is not so restricted as in the dS/dS case.  This implies that there might well be other trajectories which cross the horizon from the other hemisphere.  Using Euclidean arguments \cite{Coleman:1980aw} it was suggested that the interior region had an open de Sitter geometry.  Even though it can be argued in the dS/dS case that this holds naturally since such foliation never crosses the horizon and thus the wall remains outside the lightcone, we see that when including the mass this argument does not hold.  From Lorentzian arguments there is no preferred foliation as the whole computation only assumes $O(3)$ symmetry (see also~\cite{Cespedes:2020xpn, Cespedes:2021oso}).  Thus in order to determine the curvature of the nucleated spacetime one would have to work out the consequences having a wall,  for instance,  the effects of a wall induced anisotropy.

A possible probe of the wall would be to compute the maximum correlation function that lies within the interior de Sitter spacetime.  The correlation function on de Sitter for a massless field evaluated at equal times but at two  different points $(\rho_1,\Omega_1,\rho_2,\Omega_2)$  is given by~\cite{Allen:1987tz},
\begin{align}
G(\rho_1,\Omega_1,\rho_2,\Omega_2)\sim H^2\left(\frac{1}{1-Z}-\log(1-Z)\right) \,,
\label{correlation_function}
\end{align}
where $Z$ is the geodesic length between two points,  $Z=H^{2}\eta_{ab}X^a(r,\Omega)X^b(r',\Omega')$.  Since we are interested in correlation functions with $O(3)$ symmetry we have that $\Omega=\Omega'$,  and we get that
\begin{align}
1-Z=\sec(T)^2(1-\cos(r-r')) \,.
\end{align}
From this we see that the log dominates over the first term in the correlation function. 
Now, when the wall lies behind the lightcone the maximum value of $1-Z$  is when $\rho-\rho'=\pi/2$ in which case the correlation function becomes,
\begin{align}
G(r)\sim H^2\log(\sec(T))\sim H^3 t \,,
\end{align}
where in the last line we have used that $\sec(T)=\cosh(H t)\sim e^{H t}$.  Of course this is the usual covariance of a massless field that grows linearly with time until the end of inflation $t_e$.  Now in the presence of the wall,  if we write $r-r'=\pi/2-\Delta r$ for $\alpha$ small we get that the correlation function behaves as,
\begin{align}
G(r)\sim H^2\log(\sec(T)(1-\cos r))\sim H^2(H t+\Delta r/2+\mathcal{O}(\alpha^2)) \,,
\end{align}
which means  however that any effect of the wall is washed away by the expansion.  Still, since inflation is not eternal there is a lower bound on which scales can be access by an observer without noticing the wall.  For instance the largest distance on the CMB can be estimated from  the comoving distance to the CMB dipole to be $\chi_L\sim.46$.  This would put a lower bound on the maximum size of $\rho$ without detecting a wall.  Notice that there could still be effects from higher order correlation functions and/or on the analytical properties of the correlation function that we leave for future study.

On the other end we have the case when the false vacuum $\vert H_0\vert\to\infty$. This, we have identified as  having the same transition rate that Hartle-Hawking (or Vilenkin-Linde) in Eq.~\eqref{eq:nothingrate}.  We can see from Eq.~\eqref{eq:R*} that this limit corresponds to the turning point radius  $R^*\to 0$.  This happens because when $\vert H_0\vert$ grows the potential  moves towards the origin becoming  and narrower around $R=0$ so that the barrier becomes infinitely thin in the limit. This explains why there is no exponential suppression in this limit.

\begin{figure}[h!] 
\begin{center} 
\includegraphics[scale=0.55,trim=2.1cm 8cm 1cm 5cm,clip]{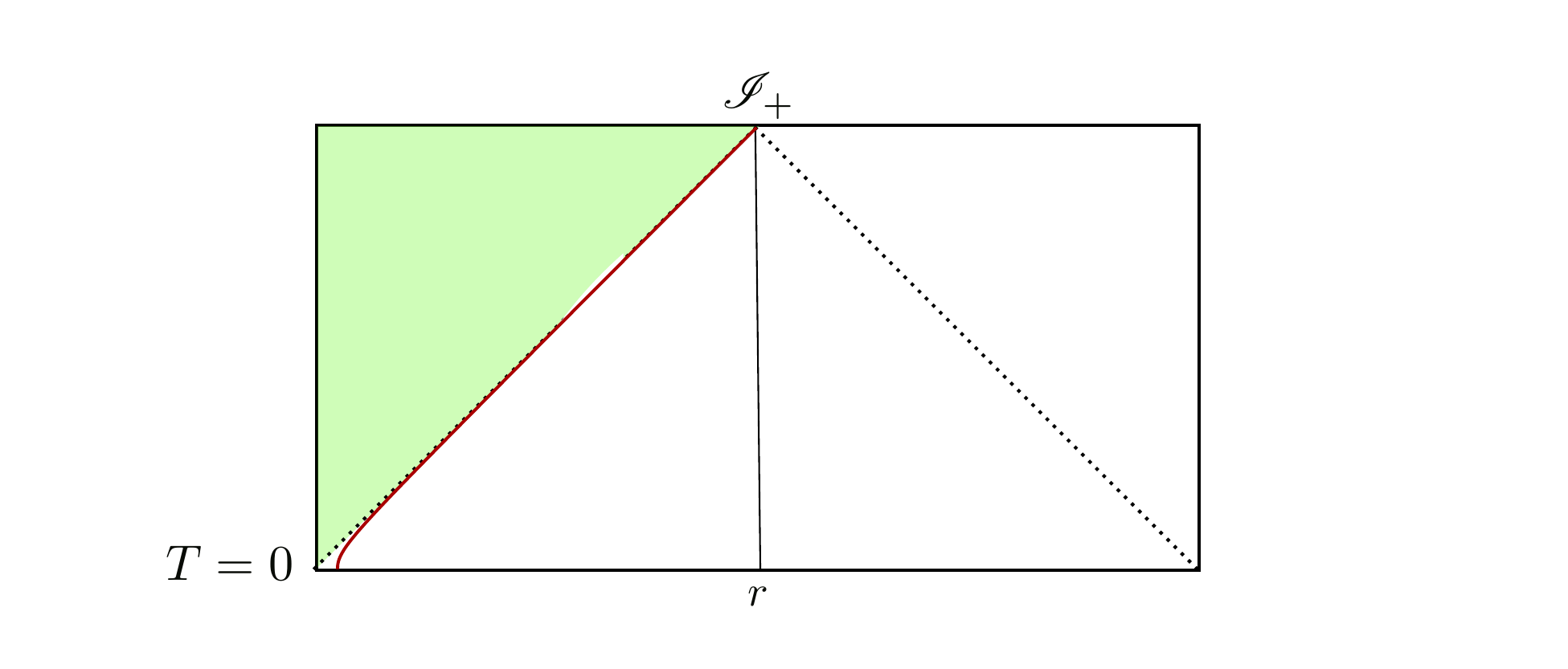}
\caption{\footnotesize{Penrose diagram for the uptunneling solution when $\vert H_0\vert\to\infty$.  The light green region is the spacetime nucleated .  The red line is the boundary of the space.  Outside the red line there is nothing.\ } \label{fig:nothing}}
\end{center} 
\end{figure}

We mentioned before that the up-tunneling transition rate is similar to tunneling from nothing.  However,  from the previous discussion we can see that there is a key difference.  As indicated in Fig.~\ref{fig:nothing} the  nucleated spacetime is not the whole de Sitter but only  a portion that includes  the whole  causal lightcone of an observer on the hemisphere of the true vacuum. There is also another small region outside the lightcone where the spacetime ends.  This starts at $r=M_{\mathrm{UV}}$ where $r$ is the radial direction in global coordinates and $M_{\mathrm{UV}}$ is the cut-off of the theory.  The reason why the spacetime starts  there is because we are still demanding the spacetime to be a solution of the junction conditions, then in the limit we are considering, the trajectory has to satisfy the geodesic equation $\cos(r)=\sqrt{1-H_0^2 R_0^2}\cos T$.  Notice that when $R_0=0$ the equations are singular. This is addressed by noticing that  $R_0$ becomes asymptotically small but  we still require that $R_0\geq 1/M_{\mathrm{UV}} $ in order for the EFT to still be valid.

This solution is actually similar to the Hawking-Turok instanton~\cite{Hawking:1998bn},  which is an Euclidean solution that gives rise to  an open universe (see however \cite{Bousso:1998ed}),  and whose wavefunction is the same as in Hartle-Hawking\footnote{The fact that up-tunneling from an infinite AdS potential corresponds to the Turok-Hawking instanton has been discussed before, \textit{eg.}\cite{Garriga:1999xh, Brown:2011gt}, although \cite{Brown:2011gt} concluded that this up-tunneling transition was not possible unlike our findings here.}.\\

Notice however that our solutions differ.  Our method is Lorentzian and then we do not need to consider singular instantons, moreover the solution can also be interpreted as the  Vilenkin-Linde wave function, although in that case, as opposed to Hartle-Hawking,  it does not satisfy detailed balance.
\section{Conclusions}

In this paper we have continued our program (see \cite{DeAlwis:2019rxg},  \cite{Cespedes:2020xpn} and \cite{Pasquarella:2022ibb}) of reformulating vacuum transitions in the Hamiltonian framework generalising the seminal work of FMP~\cite{Fischler:1990pk}. FMP considered Schwarzschild to dS transitions in the limit $M\rightarrow 0$. The dS to dS transition was studied in detail in \cite{Bachlechner:2016mtp},\cite{DeAlwis:2019rxg}. Here we have extended these works by including explicit expressions for the transition rates in all values and signs of the cosmological constants $\Lambda_A, \Lambda_B$  and all values of the black hole mass $M$. Notice that the cosmological constants are input parameters of the theory whereas the black hole mass appears as an integration constant which should be included in gravitational configurations with  spherical symmetry. 
\vskip 1mm

In general we find that for generic values of these parameters the transition rates are non-vanishing, including up-tunneling  from AdS to dS spacetimes. These non-vanishing transition rates can give rise to a huge network of vacua connected by quantum transitions among each other. This is  relevant to understand the structure and population of the string landscape.
\vskip 1mm

In the general case, the contributions to the transition rate coming from the wall is only written in terms of integrals that are only possible to solve analytically in special limits (like $M\rightarrow 0$). However the ratio of up to down tunneling is such that these contributions cancel and it gives a very simple result. This ratio allows us to verify that detailed balance is satisfied in many of the cases. In particular, due to the nature of the Hamiltonian prescription, the transition rates do not determine the sign $\eta=\pm 1$. By comparison with the standard wave function of the universe, we claim that this sign arbitrariness corresponds to the same sign arbitrariness that differentiates the Hartle-Hawking and Vilenkin probability amplitudes. In this sense we find detailed balance to work for the Hartle-Hawking but not Vilenkin. This may not be surprising since the initial conditions for the Vilenkin wave function select the expanding universe brach of the wave function of the universe which is clearly not and equilibrium situation so it's no surprise that detailed balance is not satisfied.
\vskip 1mm

Having computed the general transition rates we were able to  consider  different limits for $\Lambda_A, \Lambda_B$ and $M$. We find some interesting results. In particular for studying a transition from Minkowski spacetime we may describe Minkowski starting from several directions. First by setting $\Lambda_A=0$ to start with and take the decay rate in the $M\rightarrow 0$ limit as FMP did. This gives the finite transition rate obtained in FMP. Second, we may start with $M=0$ and consider the dS to dS transition and then take the limit $\Lambda_A\rightarrow 0$. This gives a zero transition rate $\Gamma\sim e^{-3/\Lambda_A}\rightarrow 0$. Third, we can start with the AdS to dS transition and take the limit $\Lambda_A\rightarrow 0$ which in this case reproduces the non-vanishing FMP result which is reassuring. 
\vskip 1mm

We argued that the $\Lambda_A\rightarrow 0$ limit is a more reliable definition of the Minkowski vacuum, if we start from AdS rather than from dS. In some sense AdS is  closer to Minkowski  since both are non-compact and  horizonless. Actually the SdS to SdS transitions require a more detailed analysis as discussed in the appendix, precisely because of the richer horizon structure. We offered several explanations for the difference among the limits. One is related to entropy: the dS entropy still needs to be better understood but it may be interpreted in terms of the fact that from the perspective of an observer dS is not a pure state since there are degrees of freedom that are inaccessible to the observer. On taking the $\Lambda_A\rightarrow 0$ limit the horizon moves to the surface at infinity. So it seems that  taking the limit the two entangled static patches of dS  then give rise to two disconnected but entangled Minkowski spacetimes and therefore the apparent infinite entropy. Clearly this subject needs a better understanding.
\vskip 1mm

Another interesting limit is the $\Lambda_A\rightarrow -\infty $. In this case our general expression for AdS to dS gives us a transition rate that is exactly the Hartle-Hawking and Vilenkin probabilities for the creation of dS from nothing. This result fits nicely with the Brown-Dahlen proposal for understanding nothing (of the bubble of nothing) in terms of this AdS limit. However they had concluded that the transition was not allowed and therefore questioned the Hartle-Hawking and Vilenkin results. Here we actually find perfect agreement between two independent calculations and provide then a strong case for this interpretation. We find this result remarkable. In general the limit $\Lambda_A\rightarrow -\infty $ should not be trusted since it brings us beyond the validity of effective field theory. The proper limit to take is to compare $\Lambda_A$ with the other scales of the system, like the brane tension $\kappa$ and the dS cosmological constant $\Lambda_B$ and consider the regime $|\Lambda_A| \gg  \Lambda_B, \kappa^2, M^2$. In this case we obtain again the Hartle-Hawking/Vilenkin probability at leading order plus corrections of order
$\Lambda_B/|\Lambda_A \ll 1$, $ M^2/|\Lambda_A| \ll 1$ and $\kappa^2/|\Lambda_A| \ll 1$. 
\vskip 1mm 

Note, however that our identification of the nucleated spacetime fits with the Hawking-Turok instanton (and its open universe interpretation) rather than the creation of a full closed spacetime as in the Hartle-Hawking and Vilenkin cases. Since the transition rate in both cases is the same, we may conjecture that also for these cases there is a non-vanishing probability to create AdS and Minkowski spacetimes if we assign vanishing entropies to these spaces. Clearly this needs to be better understood.
\vskip 1mm

Finally we may consider the regime in which the black hole mass dominates ($M\gg \Lambda_A, \Lambda_B$). This is a much less understood limit that may need further study. In particular, unlike the small mass limit, the ratio of up and down transition rates does not give exponential of the entropies. Instead of entropies we find differences of areas that do not correspond to horizon areas. It may be tempting to speculate these are some generalised entropies but we do not yet have a proper interpretation.
\vskip 1mm

Regarding the trajectory of the wall for the generic $M\neq 0$ case, we find it interesting that it does not correspond to a geodesic that favours the open slicing of dS, contrary to the $M=0$ case. This leaves open the question if an open universe is a general implication of the bubble nucleation process.
\vskip 1mm
Some approaches to the address the cosmological constant problem envisage the landscape transitions to Minkowski or near Minkowski (with positive but tiny cosmological constants) and with AdS being a terminal  spacetime  (for recent discussions see \cite{Giudice:2021viw,Kartvelishvili:2020thd,Friedrich:2022tqk}).  It would be interesting to explore what will be the implications that up-tunellling from AdS could have to those proposals.
\\
There are many questions still open. Extending the Hamiltonian approach to include scalar field potentials is a totally open question. An AdS/CFT interpretation of our results may be interesting, following the lines of  \cite{Maldacena:2010un} that interpreted the CDL Minkowski to AdS transition holographically. Furthermore \cite{Pasquarella:2022ibb} proposed interesting holographic interpretations for the 2D transitions. Extensions of these ideas to the 4D systems considered here may be worth exploring.
\vskip 1mm

Clearly our results leave  plenty of issues still to be understood. But we hope that our work and the questions we raised may guide future progress in this area.

\section*{Acknowledgements}
We enjoyed interesting conversations on these topics with Cliff Burgess, Sergei Dubovsky, Steve Gratton, Ted Jacobsen, Veronica Pasquarella, Suvrat Raju, Jorge Santos, Ashoke Sen, Lenny Susskind and Aron Wall.  The work of FQ has been partially supported by STFC consolidated grants ST/P000681/1, ST/T000694/1. FM is funded by a UKRI/EPSRC Stephen Hawking fellowship, grant reference EP/T017279/1, partially supported by the STFC consolidated grant ST/P000681/1 and funded by a G- Research grant for postdocs in quantitative fields.  The work of SC is  supported by STFC consolidated grant ST/T000791/1. SdA and FQ thank Perimeter Institute for hospitality where part of this project developed.

\appendix
\label{sec:Appendix}

\section{Schwarzschild black hole to de Sitter\label{sec:stods}}

In this section we will give the details of the bulk term FMP calculation and  extend it to the SdS case. FMP discuss the case studied by BGG - FGG\citep{Blau:1986cw,Farhi:1989yr}, i.e. the nucleation of a dS
space by tunneling from a Schwarzschild black hole. In this case, $A_{I}=1-2GM/R$ and $A_{O}=1-H^{2}R^{2}$. As in the discussion
after equation \eqref{eq:EofM}, at the turning point $V=-1$ $\hat{A}_{O}>0,\hat{A}_{I}>0$, but there are now two turning points $R_{1}<R_{2}$ such that $R_{{\rm b}}<R_{1,2}<R_{{\rm c}}$, where $R_{{\rm b}}=2GM$ is the black hole horizon (in this case the same as the Schwarzschild) while $R_{{\rm D}}=H^{-1}$ is the cosmological horizon (in this case the same as the dS).\\

$\hat R'_{+}$ vanishes at $\hat{R}^{3}=\frac{R_{{\rm b}}}{H^{2}+\kappa^{2}}\equiv R_{<}^{3}$
and $\hat R'_{-}$ vanishes at $\hat{R}^{3}=\frac{R_{{\rm b}}}{H^{2}-\kappa^{2}}\equiv R_{>}^{3}$. Note that if $M$ (which is an integration constant) is such that $M = M_{{\rm b}}\equiv\frac{1}{2G\sqrt{H^{2}+\kappa^{2}}}$ the turning point $R_{1}=R_{<}=R_{{\rm b}}$ and if $M=M_{{\rm c}}=\frac{H^{2}-\kappa^{2}}{2GH^{3}}$
(note that $M_{{\rm b}}>M_{{\rm c}}$), the turning point $R_{2}=R_{>}=R_{{\rm c}}$. 

From the matching condition it follows that $R'_{-}$ is negative
for $\hat{R}$ $>R_{>}$ i.e. to the right of the point where $R'_{-}$
vanishes while $R'_{+}$ is negative for $\hat{R}>R_{_{<}}$. Thus
we have for the bulk contribution at the turning point geometry,
\begin{align}
S_{{\rm Bu}} & =\frac{i\eta\pi}{G}\left[\int_{0}^{\hat{r}}dr\frac{dR}{dr}R \, \theta\left(-R'\right)+\int_{\hat{r}}^{r_{\rm max}}dr\frac{dR}{dr}R \, \theta\left(-R'\right)\right] = \nonumber \\
 & =\frac{i\eta\pi}{2G}\left[\left(\hat{R}^{2}-R_{{\rm D}}^{2}\right)\,\theta\left(\hat{R}-R_{>}\right)+\left(R_{{\rm b}}^{2}-\hat{R}^{2}\right)\,\theta\left(\hat{R}-R_{<}\right)\right] \,.\label{eq:SbtodS}
\end{align}

The wall contribution cannot be calculated analytically and FMP did
not do so. All that matters for our present purposes is that it is
finite. We also note that in the limit $M\rightarrow0$ it was calculated in \citep{Bachlechner:2016mtp,DeAlwis:2019rxg}.

\begin{figure}[h!] 
\begin{center} 
\includegraphics[scale=0.5, trim = 1cm 2cm 0cm 1cm, clip]{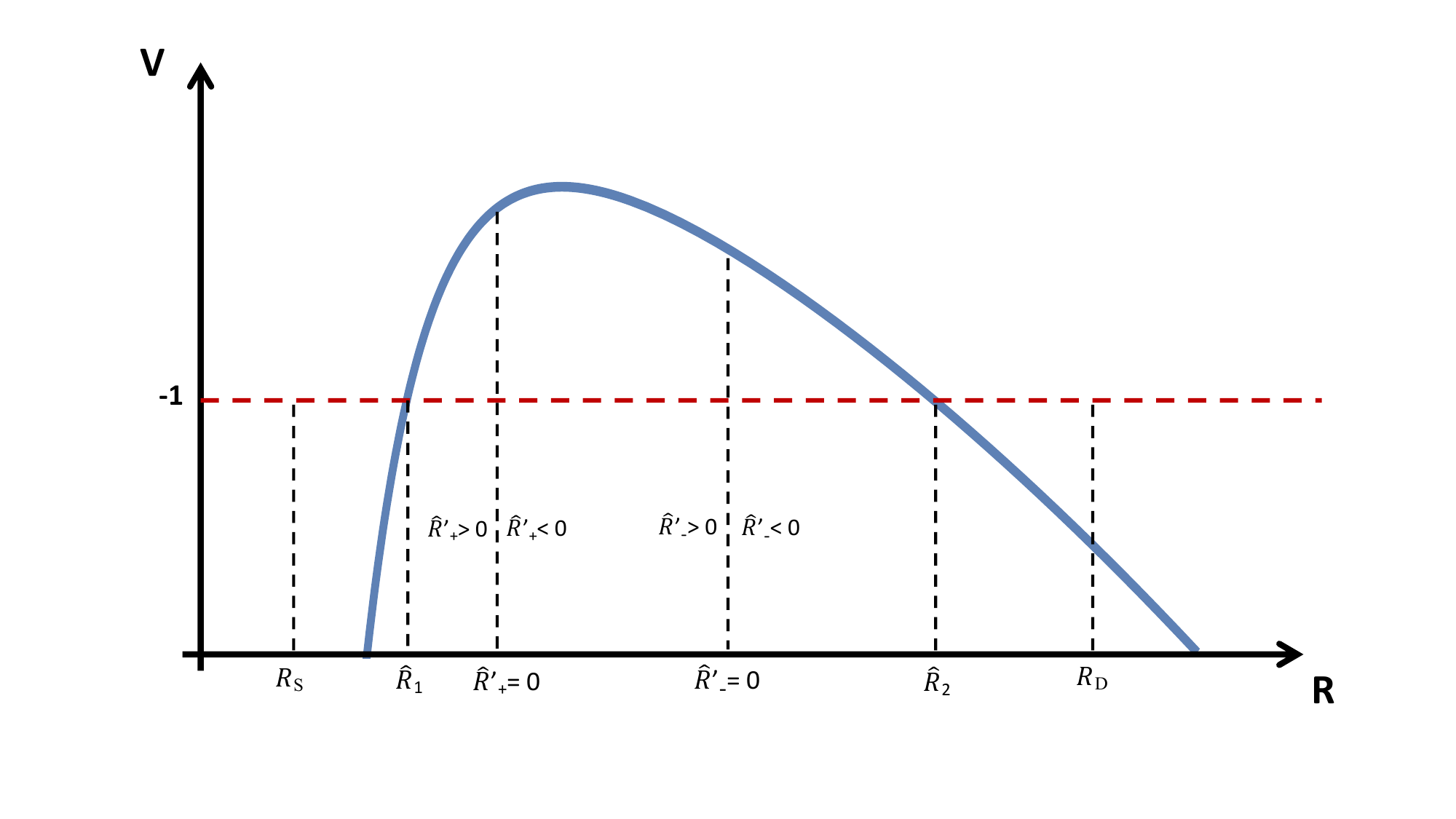}
\caption{\footnotesize{.} \label{fig:potentialdS}}
\end{center} 
\end{figure}

There are three cases to consider: 
\begin{itemize}[leftmargin=*]
\item[]{Case \textit{i)}}\\
For large black hole mass $M>M_{{\rm b}}>M_{c}$, $R_{<}<R_{1}<R_{2}<R_{>}$
we have $S_{{\rm Bu}}(R_{2})=\frac{i\eta\pi}{2G}\left(R_{{\rm b}}^{2}-R_{2}^{2}\right)$
since $R_{2}>R_{<}$ and $S_{{\rm Bu}}(R_{1})=\frac{i\eta\pi}{2G}\left(R_{{\rm b}}^{2}-R_{1}^{2}\right)$.
Hence
\be
\label{eq:FMP1}
S_{{\rm Bu}}\left(R_{2}\right)-S_{{\rm Bu}}(R_{1})=-\frac{i\eta\pi}{G}\left(R_{{\rm 2}}^{2}-R_{1}^{2}\right)\,.
\ee

\item[]{Case \textit{ii)}}\\
The second case is for $M_{{\rm b}}>M>M_{{\rm c}}$. Now $R_{1}<R_{<}<R_{2}<R_{>}$
and we get $S_{B}(R_{2})=\frac{i\eta\pi}{2G}\left(R_{{\rm b}}^{2}-R_{2}^{2}\right)$
and $S_{B}(R_{1})=0$. Hence
\be
\label{eq:FMP2}
S_{{\rm Bu}}\left(R_{2}\right)-S_{{\rm Bu}}(R_{1})=-\frac{i\eta\pi}{G}\left(R_{{\rm 2}}^{2}-R_{{\rm b}}^{2}\right) \,.
\ee

\item[]{Case \textit{iii)}}\\
The third case is for small black holes $M<M_{{\rm c}}<M_{{\rm b}}$
so that $R_{{\rm b}}<R_{1}<R_{2}<R_{{\rm D}}$. Here we have $S_{{\rm Bu}}(R_{2})=\frac{i\eta\pi}{2G}\left(R_{{\rm b}}^{2}-R_{{\rm D}}^{2}\right)$
and $S_{B}(R_{1})=0$ so that
\be
\label{eq:FMP3}
S_{{\rm Bu}}\left(R_{2}\right)-S_{{\rm Bu}}(R_{1})=-\frac{i\eta\pi}{G}\left(R_{{\rm {\rm D}}}^{2}-R_{{\rm b}}^{2}\right) \,.
\ee
\end{itemize}

Thus we have given,  for the reader's convenience the derivation of Eq.~(48) of FMP.

\subsubsection{de Sitter to Schwarzschild black hole}

In the down-tunneling from dS to a Schwarzschild black hole, $A_{I}=1-2GM/R$ and $A_{O}=1-H^{2}R^{2}$. The matching condition gives
\begin{equation}
\frac{\hat{R}'_{\pm}}{L}=\frac{1}{2\kappa\hat{R}}\left[-2GM+\left(H^{2}\mp\kappa^{2}\right)\hat{R}^{3}\right]\,.\label{eq:R'dStoBH}
\end{equation}

Note that if $M=0$ $\hat{R}'_{-}>0$ for all $R$. If one goes back
for a moment to the dS to dS transitions for the bulk integral contribution
we had a term $\theta\left(-R'_{-}\right)H_{I}^{-2}$, and in this
case since we have Minkowski inside $H_{I}\rightarrow0$ and this
term is potentially divergent. However as we have just observed $\hat{R}'_{-}>0$
so that $\theta\left(-R'_{-}\right)=0$ and the potential divergence is avoided. This is why one gets a finite answer for down tunneling from dS to Minkowski even though there is no horizon to cutoff the integral in Minkowski. 

However in the present case as we see from Eq.~\eqref{eq:R'dStoBH} there
is a regime, namely $\hat{R}^{3}<\frac{2GM}{H^{2}+\kappa^{2}}\equiv R_{<}^{3}$ where $\theta\left(-\hat{R}'_{-}\right)=1$. So let us consider the same three cases that we analyzed in the previous subsection. It should be noted now that the region where $\hat R'_{\pm }$ is negative is to the right of (respectively) $R_<$ and $R_>$. 

So for general $\hat R $ Inside the barrier, we have for the bulk integral

\begin{equation}
S_{{\rm Bu}}^{{\rm down}}(\hat R)=\frac{i\eta\pi}{2G}[(\hat R^2-R^2_{c})\theta (R_<-\hat R)+(R^2(r_{{\rm max}})-\hat R^2)\theta (R_>-\hat R)].\label{eq:BHtodS}
\end{equation}
Note that here we have introduced a cutoff in the first term $R_c$ in the first term  (which corresponds to the black hole region $r<\hat r$) which would have been there in the SdS case but for FMP case which corresponds to a asymptotically Minkowski black hole, must be sent to infinity. This however will make this transition ill-defined in the last two caes as  will see below.
For case $i)$, i.e. $M>M_{{\rm b}}>M_{c}$, $R_{<}<R_{1}<R_{2}<R_{>}$ we have $S_{{\rm Bu}}^{{\rm down}}(R_{2})=\frac{i\eta\pi}{2G}\left(R^{2}(r_{\rm max})-R_{2}^{2}\right)$ and $S_{{\rm Bu}}^{{\rm down}}\left(R_{1}\right)=\frac{i\eta\pi}{2G}\left(R^{2}(r_{\rm max})-R_{1}^{2}\right)$.
So the total contribution,\footnote{As before we just focus on one value of $\eta$ in the numerator and denominator of the ratio of wave functions.} is 
\be
S_{{\rm Total}}^{{\rm down}}\left(R_{2}-R_{1}\right)=-\frac{i\eta\pi}{2G}\left(R_{2}^{2}-R_{1}^{2}\right)+S_{W}\left(R_{2}-R_{1}\right) \,,
\ee

which is exactly the same as the total up-tunneling action
difference, see Eq.~\eqref{eq:FMP1}. This implies that the up and down transition rates from large black holes to dS is the same in the leading WKB approximation. In the other two cases
however we have an ill defined results for down-tunneling.

In case $ii)$, i.e. $M_{{\rm b}}>M>M_{{\rm c}},\,R_{1}<R_{<}<R_{2}<R_{>}$, we have $S_{{\rm Bu}}^{{\rm down}}(R_{2})=\frac{i\eta\pi}{2G}\left(R^{2}(r_{\rm max})-R_{2}^{2}\right)$
and $S_{{\rm Bu}}^{{\rm down}}(R_{1})=\frac{i\eta\pi}{2G}\left(-R^{2}_c+R^{2}(r_{\rm max})\right)$.
Hence we have 
\be
\label{eq:down2}
S_{{\rm Total}}^{{\rm down}}\left(R_{2}-R_{1}\right)=-\frac{i\eta\pi}{2G}\left(R_{2}^{2}-R^{2}_c\right)+S_{W}\left(R_{2}-R_{1}\right) \,.
\ee

Finally for case $iii)$, i.e. $M<M_{{\rm c}}<M_{{\rm b}},\,R_{1}<R_{<}<R_{>}<R_{2}$ we have
$S_{{\rm Bu}}^{{\rm down}}(R_{2})=0$ and $S_{{\rm Bu}}^{{\rm down}}(R_{1})=\frac{i\eta\pi}{2G}\left(-R^{2}_c)\right)$.
So the total down tunneling amplitude is
\be
\label{eq:down3}
S_{{\rm Total}}^{{\rm down}}\left(R_{2}-R_{1}\right)=-\frac{i\eta\pi}{2G}\left(-R^{2}_c\right)+S_{W}\left(R_{2}-R_{1}\right) \,.
\ee

The classical actions in these last two down-tunneling cases is ill-defined since for a Minkowskian black hole $R_c\rightarrow \infty$. The problem in Eq.~\eqref{eq:down2} and \eqref{eq:down3} lies in the fact that in cases $ii)$ and $iii)$ the junction condition in Eq.~\eqref{eq:R'dStoBH} dictates that $R' < 0$ in the spacetime inside the bubble. Hence $R_c$ corresponds to the maximum value of $R$ for the Schwarzschild spacetime, which is infinity since there is no cosmologial horizon.\footnote{Note that $R(r_{\rm max})$ correspond to the minimum value (as $R' < 0$) of $R$ for the dS spacetime,
which is $0$.} What is well-defined for both up and down transitions involving black hole space times are transitions between SdS spaces. This we discuss in the next subsection.

\subsection{Schwarzschild-de Sitter to Schwarzschild-de Sitter\label{sec:sdstosds}}

Finally we observe that in none of these cases is detailed balance
satisfied. Next we will follow these arguments to derive the corresponding equations for the rather more complicated case of transitions from SdS to SdS. In the general case,
\begin{equation}
A=1-\frac{2GM}{R}-H^{2}R^{2}=-\frac{H^{2}}{R}\left(R-R_{-}\right)\left(R-R_{{\rm b}}\right)\left(R-R_{c}\right).\label{eq:SdS-A}
\end{equation}
The parameters are taken to be such that $3\sqrt{3}GM<H^{-1}$ in
which case there is one negative real root and two positive real roots
which satisfy
\begin{equation}
R_{-}<0<2GM<R_{{\rm b}}<3GM<R_{c} \,.\label{eq:SdS-roots}
\end{equation}
The smaller positive root $R_{b}$ is identified with the horizon of the black hole and the larger one $R_{c}$ with the cosmological
 horizon. We consider transitions from one SdS to another with
parameters inside the bubble being denoted with the subscript $I$
and those outside with $O$. Writing $\Delta M\equiv M_{O}-M_{I}$
and $\Delta H^{2}\equiv H_{O}^{2}-H_{I}^{2}$, we have the matching
condition
\begin{equation}
\frac{\hat{R}_{\pm}}{L}=\frac{1}{2\kappa\hat{R}}\left[2G\Delta M+\left(\Delta H^{2}\mp\kappa^{2}\right)\hat{R}^{3}\right].\label{eq:SdS-match}
\end{equation}
Let us simplify the discussion by choosing $\kappa^{2}<|\Delta H^{2}|$
so that the sign of the second term in square brackets is determined
by the sign of $\Delta H^{2}$. Now as in the FMP/BGG discussion the
sign of $\hat{R}_{\pm}$ changes at a zero but unlike in that case
if $\Delta M$ and $\Delta H^{2}$ have the same sign then $\hat{R}_{\pm}$
does not vanish anywhere. To remain close to the FMP discussion let
us first choose these signs to be opposite i.e. $\epsilon(\Delta M)=-\epsilon(\Delta H^{2})$
with $\Delta M>0$\footnote{This choice goes over to the FMP case when $M_{I}\rightarrow0$ and $H_{O}\rightarrow0$.} so that $\hat{R}_{+}$ vanishes at $R_{<}$ and $\hat{R}_{-}$ vanishes at $R_{>}$ where 
\begin{equation}
R_{<}^{3}=\frac{2G\Delta M}{|\Delta H^{2}|+\kappa^{2}} \,,\qquad R_{>}^{3}=\frac{2G\Delta M}{|\Delta H^{2}|-\kappa^{2}} \,.\label{eq:R<>}
\end{equation}
These points coincide with the turning points $R_{1.2}$ when the
masses (which we recall are integration constants), are such that
$R_{<}=R_{{\rm b}}$ and $R_{>}=R_{c}$.

\begin{figure}[h!] 
\begin{center} 
\includegraphics[scale=0.5, trim = 1cm 2cm 0cm 1cm, clip]{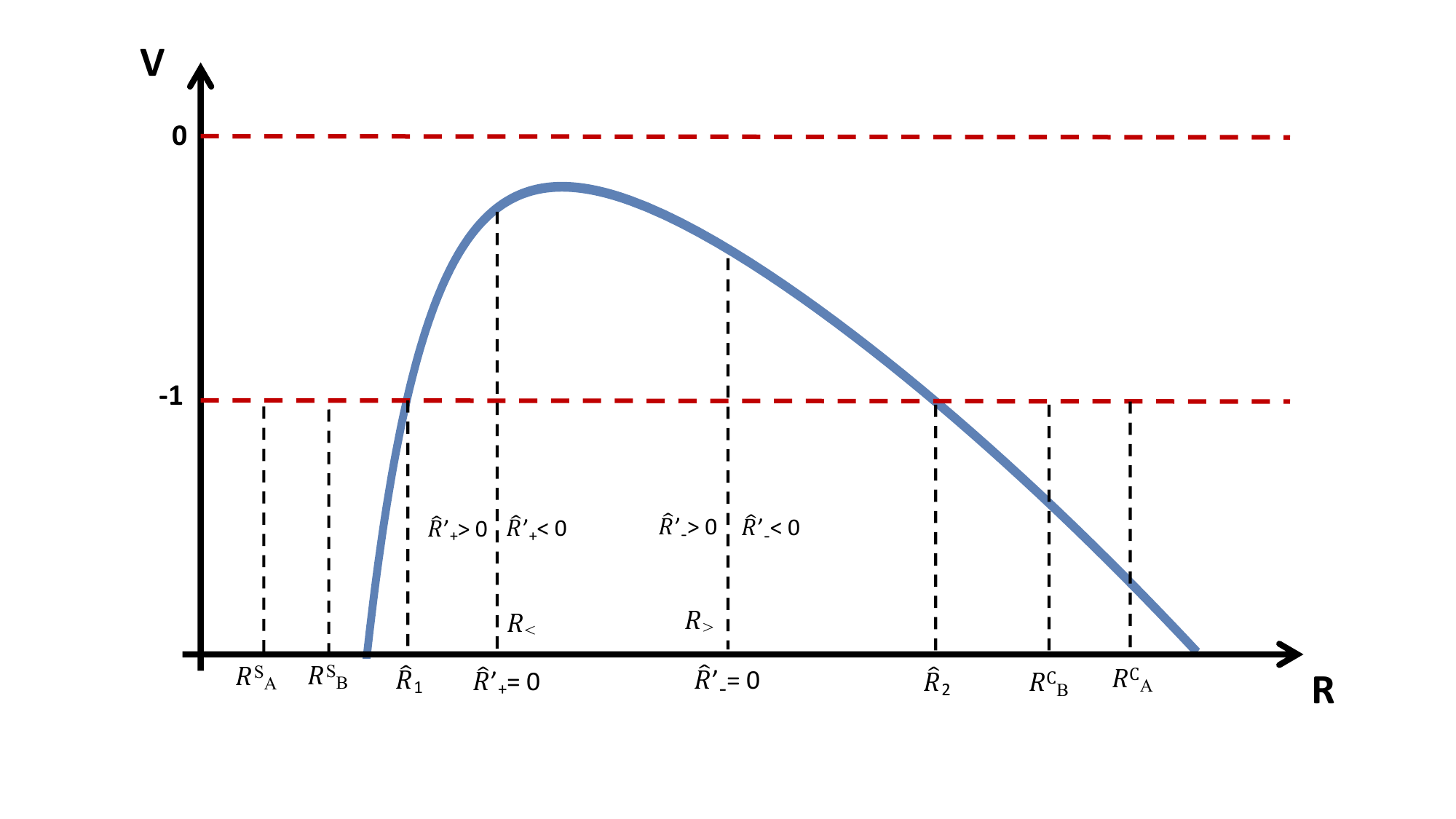}
\caption{\footnotesize{.} \label{fig:potentialSdS}}
\end{center} 
\end{figure}

So let us consider the tunneling $A\rightarrow B$ so that we put
${\rm O=A}$ and ${\rm I}=B$. We will be interested mainly in what
corresponds to case $iii)$ of FMP i.e. the small black hole case\footnote{This is because ultimately we would like to send the black masses to zero.}. Then we have $R_{1}<R_{<}<R_{>}<R_{2}$. For the bulk action we
have (see first line of eqn. \eqref{eq:SbtodS} and figure \eqref{fig:potentialSdS})
\begin{equation}
S_{{\rm Bu}}(\hat{R})=\frac{i\eta\pi}{2G}\left[\left(\hat{R}^{2}-\left(R_{I,c}\right)^{2}\right)\theta\left(\hat{R}-R_{>}\right)+\left(\left(R_{{\rm O,{ b}}}\right)^{2}-\hat{R}^{2}\right)\theta\left(\hat{R}-R_{<}\right)\right].\label{eq:SdStoSdS}
\end{equation}
So at the turning points we have
\begin{align*}
S_{{\rm Bu}}(\hat{R_{2}}) & =\frac{i\eta\pi}{2G}\left[\left(\hat{R}_{2}^{2}-\left(R_{c}^{B}\right)^{2}\right)\theta\left(\hat{R}_{2}-R_{>}\right)+\left(\left(R_{{\rm {\rm b}}}^{A}\right)^{2}-\hat{R}_{2}^{2}\right)\theta\left(\hat{R}_{2}-R_{<}\right)\right] =\\
 & =\frac{i\eta\pi}{2G}\left[-\left(R_{c}^{B}\right)+\left(R_{{\rm {\rm b}}}^{A}\right)^{2}\right],\\
S_{{\rm Bu}}(\hat{R_{1}}) & =\frac{i\eta\pi}{2G}\left[\left(\hat{R}_{1}^{2}-\left(R_{c}^{B}\right)^{2}\right)\theta\left(\hat{R}_{1}-R_{>}\right)+\left(\left(R_{{\rm {\rm b}}}^{A}\right)^{2}-\hat{R}_{1}^{2}\right)\theta\left(\hat{R}_{1}-R_{<}\right)\right] =\\
 & =0 \,.
\end{align*}
Defining $F\left[R_{2}-R_{1}\right]\equiv F\left[R_{2}\right]-F\left[R_{1}\right]$
we have for $i)$ times the total action (including the wall term)
\begin{equation}
I^{AB}\left[R_{2}-R_{1}\right]=\frac{\eta\pi}{2G}\left[\left(R_{c}^{B}\right)^{2}-\left(R_{{\rm {\rm b}}}^{A}\right)^{2}\right]+I_{W}^{AB}\left[R_{2}-R_{1}\right].\label{eq:IAB}
\end{equation}
Note that this is the same as FMP case $iii)$. Interchanging $A$ and
$B$ we have
\be
I^{BA}\left[R_{2}-R_{1}\right]=\frac{\eta\pi}{2G}\left[\left(R_{c}^{A}\right)^{2}-\left(R_{{\rm {\rm b}}}^{B}\right)^{2}\right]+I_{W}^{BA}\left[R_{2}-R_{1}\right] \,.
\ee

The wall term is clearly symmetric under the interchange i.e. $I_{W}^{AB}=I_{W}^{BA}$. Hence we have for the ratio of the probabilities of going from $A$ to $B$ to the reverse,
\begin{equation}
\frac{P^{AB}}{P^{BA}}=\frac{\exp\left\{ \frac{\eta\pi}{G}\left[\left(R_{c}^{B}\right)^{2}-\left(R_{{\rm {\rm b}}}^{A}\right)^{2}\right]+2I_{W}^{AB}\left[R_{2}-R_{1}\right]\right\} }{\exp\left\{ \frac{\eta\pi}{G}\left[\left(R_{c}^{A}\right)^{2}-\left(R_{{\rm {\rm b}}}^{B}\right)^{2}\right]+2I_{W}^{BA}\left[R_{2}-R_{1}\right]\right\} }=e^{S^{B}-S^{A}}.\label{eq:detailedbalance}
\end{equation}
Here we have defined $S^{B}=\frac{\eta\pi}{G}\left[\left(R_{c}^{B}\right)^{2}+\left(R_{{\rm {\rm b}}}^{B}\right)^{2}\right]$
and $S^{A}=\frac{\eta\pi}{G}\left[\left(R_{c}^{A}\right)^{2}+\left(R_{{\rm {\rm b}}}^{A}\right)^{2}\right]$
which are the sum of the horizon entropies of each SdS space. If 
we can interpret this sum as the total entropy of SdS space (we know
of no direct demonstration of this) then indeed the relation Eq.~\eqref{eq:detailedbalance}
is the statement of detailed balance.

\subsubsection*{Graphic display of the computation}

We can give an intuitive understanding of the previous computations as described in this subsection.

First, note that the constraint in Eq.~\eqref{eq:TurningPoints} dictates the sign of $R'$, that can only changes at the horizons, i.e. where $R' = 0$. In a compound state with two spacetimes separated by a wall at $\hat{r}$, the sign of $R'$ in the vicinity of the wall is determined by the junction conditions in Eq.~\eqref{eq:JunctionConditions}. As can be observed in Fig.~\ref{fig:potentialSdS}, the junction conditions imply that there are only three possible combinations of the signs of $\hat{R}'_{\pm}$, namely
\begin{equation}
R'_{\pm} = \{(+,+), (+, -), (-, -)\} \,. \label{eq:RpOptions}
\end{equation}

\begin{figure}
\centering
\begin{minipage}{.5\textwidth}
  \centering
  \includegraphics[width=\linewidth]{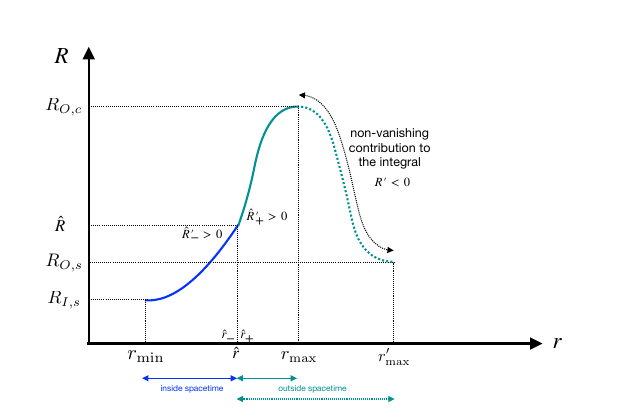}
  \captionof{figure}{\footnotesize{Configuration with $R'_\pm = (+,+)$.}}
  \label{fig:SdS1}
\end{minipage}%
\begin{minipage}{.5\textwidth}
  \centering
  \includegraphics[width=\linewidth]{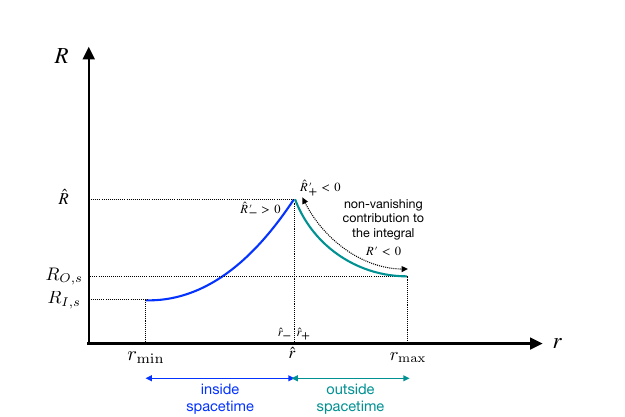}
  \captionof{figure}{\footnotesize{Configuration with $R'_\pm = (+,-)$.}}
  \label{fig:SdS2}
\end{minipage}
\end{figure}

\begin{figure}[h!] 
\begin{center} 
\includegraphics[scale=1]{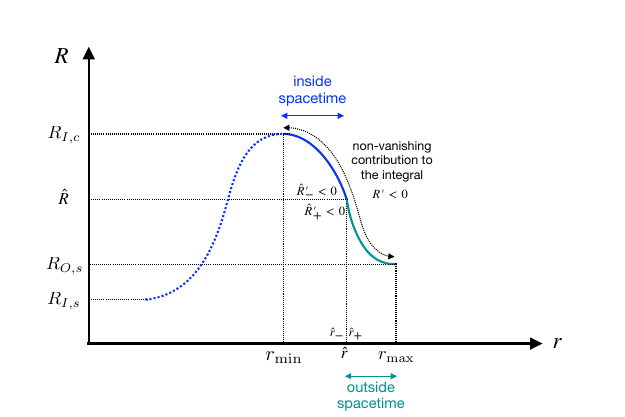}
\caption{\footnotesize{Configuration with $R'_\pm = (-,-)$.} \label{fig:SdS3}}
\end{center} 
\end{figure}

The three options in Eq.~\eqref{eq:RpOptions} can be visualized as in Fig.~\ref{fig:SdS1}, \ref{fig:SdS2}, \ref{fig:SdS3}. We consider the compound state to be given by the region of spacetime connected to the wall, i.e. it extends up to the closest horizon both in the inside and the outside spacetime regions. For instance, in Fig.~\ref{fig:SdS3}  the dotted line is not part of the spacetime, as the closest horizon to the wall in the inside region is reached the cosmological horizon at $r_{\text{min}}$.

The three cases $i)$, $ii)$ and $iii)$ in the previous section correspond to transitions between pairs of compound states like those displayed in Fig.~\ref{fig:SdS1}, \ref{fig:SdS2}, \ref{fig:SdS3}. In particular:
\begin{center}
Case $i)$ $\quad \Leftrightarrow \quad (+,-) \rightarrow (+, -)$ \\
Case $ii)$ $\quad \Leftrightarrow \quad (+,+) \rightarrow (+, -)$ \\
Case $iii)$ $\quad \Leftrightarrow \quad (+,+) \rightarrow (-, -)$
\end{center}

For each configuration, it is easy to evaluate the contribution to the bulk integral: from Eq.~\eqref{eq:Bulk} it is clear that only the regions where $R'<0$ contribute with a factor $R^2$ evaluated at the extrema of these regions. For instance, for the configuration $R'_\pm = (-, -)$ in Fig.~\ref{fig:SdS3}, the non-vanishing contribution comes from the region between the cosmological horizon of the interior spacetime at $r_{\rm min}$ and the black hole horizon of the exterior spacetime at $r_{\rm max}$. The value of the action would then be

\begin{align}
S_B &= \frac{i \eta \pi}{G} \left[\int_{r_{\rm min}}^{\hat{r}} dr R' R \, \theta(-R'_-) + \int_{\hat{r}}^{r_{\rm max}} dr R' R \, \theta(-R'_+)\right] = \\
&= \frac{i \eta \pi}{G} \left[\left(R_{O, s}^2 - \hat{R}^2\right) + \left(\hat{R}^2 - R_{O, c}^2\right)\right] = \frac{i \eta \pi}{G} \left(R_{O, s}^2 - R_{O, c}^2\right) \,.
\end{align}

Therefore, defining as in FMP
\begin{equation}
F_I[\hat{R}] = \frac{i \eta \pi}{G} \int_{R(r_{\rm min})}^{R(\hat{r})} dR R \, \theta(-R'_-) \,, \qquad F_O[\hat{R}] = \frac{i \eta \pi}{G} \int_{R(\hat{r})}^{R(r_{\rm max})} dR R \, \theta(-R'_+) \,,
\end{equation}
we can compute the bulk action case by case as in the previous section, using that
\begin{equation}
S_B \equiv F[\hat{R}_2 - \hat{R}_1] \equiv F_I[\hat{R}_2 - \hat{R}_1] + F_O[\hat{R}_2 - \hat{R}_1] \,,
\end{equation}
and $F_i[\hat{R}_2 - \hat{R}_1] = F[\hat{R}_2] - F[\hat{R}_1]$ for $i = 1,2$. Note that $\hat{R}_{1,2}$ are the turning points defined in the previous subsection.

\begin{itemize}[leftmargin=*]
\item[]{Case \textit{i)}}\\
In this case, both $\hat{R}_1$ and $\hat{R}_2$ fall in a region where $\hat{R}'_\pm = (+, -)$, corresponding to Fig.~\ref{fig:SdS2}. Then
\begin{equation}
F_I[\hat{R}_2 - \hat{R}_1] = 0 \,, \qquad F_O[\hat{R}_2 - \hat{R}_1] = \frac{i \eta \pi}{2G} \left(\hat{R}_2^2 - \hat{R}_1^2\right) \,,
\end{equation}
where $F_I$ gives zero contribution because in the interior spacetime $R'$ is always positive. This implies
\begin{equation}
S_B = \frac{i \eta \pi}{2G} \left(\hat{R}_2^2 - \hat{R}_1^2\right) \,.
\end{equation}
\item[]{Case \textit{ii)}}\\
In this case, $\hat{R}_1$ is in a region where $\hat{R}'_\pm = (+, +)$, while $\hat{R}_2$ is in a region where $\hat{R}'_\pm = (+, -)$ corresponding to Fig.~\ref{fig:SdS1} and Fig.~\ref{fig:SdS2} respectively. Then
\begin{equation}
F_I[\hat{R}_2 - \hat{R}_1] = 0 \,, \qquad F_O[\hat{R}_2 - \hat{R}_1] = \frac{i \eta \pi}{2G} \left(R_{O, b}^2 - \hat{R}_2^2\right) \,,
\end{equation}
where $F_I$ gives zero contribution because in the interior spacetime $R'$ is always positive. This implies
\begin{equation}
S_B = \frac{i \eta \pi}{2G} \left(R_{O, b}^2 - \hat{R}_2^2\right) \,.
\end{equation}

We observe that these two cases give exactly the same result as in FMP i.e. \eqref{eq:FMP1} and \eqref{eq:FMP2}.
\item[]{Case \textit{iii)}}\\
In this case $\hat{R}_1$ is in the region with $R'_{\pm} = (+,+)$, while $\hat{R}_2$ is in the region $\hat{R}_{\pm} = (-,-)$, corresponding to Fig.~\ref{fig:SdS1} and Fig.~\ref{fig:SdS3} respectively. Then
\begin{equation}
F_I[\hat{R}_2 - \hat{R}_1] = \frac{i \eta \pi}{2G}\left(\hat{R}_2^2 - R_{I, c}^2\right) \,, \qquad F_O[\hat{R}_2 - \hat{R}_1] = \frac{i \eta \pi}{2G} \left(R_{O, b}^2 - \hat{R}_2^2\right) \,,
\end{equation}
which implies
\begin{equation}
S_B = \frac{i \eta \pi}{2G} \left(R_{O, b}^2 - R_{I, c}^2\right) \,,
\end{equation}
which coincides with the result found in the last subsection. The contribution of figure Fig.~\ref{fig:SdS1} however is clearly a disconnected term of sort discussed in the next subsection and cancels between the two turning points. 
\end{itemize}

\subsection{Disconnected terms}

Here we derive a general formula for the bulk contribution to the
action for transitions from and to space times with two horizons.
The formula seems to have been first written down in \citep{Bachlechner:2016mtp}
(see eqn. (3.37)) but without a detailed derivation. Here for the
readers convenience we include these details and also highlight some
differences with Bachlechner's discussion \citep{Bachlechner:2016mtp}.

For the turning point geometries the bulk action $S_{B}$ simplifies
with the first term in square brackets in  \eqref{eq:SclassicalBulk} giving zero and the second term
contributes only when $\epsilon(R')=-1$ i.e. whenever $\cos^{-1}\left(\frac{R'}{L\sqrt{A_{{\rm I,{\rm O}}}}}\right)=\pi$.
Thus we have \footnote{We've taken the integration in $r$ all the way up to infinity even though in practice it may be cutoff at some finite value. }.
\begin{eqnarray}
iS_{{\rm Bu}}(\hat{R}_{r}) & =- & \frac{\eta\pi}{G}\left[\int_{0}^{\hat{r}}drR'R\theta(-R_{-}')+\int_{\hat{r}}^{\infty}drR'R\theta(-R_{+}')\right]\label{eq:SButp}
\end{eqnarray}
Let us focus on the second integral.
\begin{align}
\int_{\hat{r}}^{\infty}drR'R\theta(-R_{+}') & =\int_{\hat{r}}^{\infty}dr\frac{1}{2}\left(\frac{dR^{2}}{dr}\right)\theta(-R_{+}')\nonumber \\
 & =\int_{\hat{r}}^{\infty}dr\frac{1}{2}\left(\frac{d}{dr}R^{2}\theta(-R_{+}')\right)-\int_{\hat{r}}^{\infty}dr\frac{1}{2}R^{2}\frac{d}{dr}\theta(-R_{+}').\label{eq:I1-I2}
\end{align}
Let us first consider the first integral above. Observe that at the
turning points for the geometry $R'/L=\pm\sqrt{A_{O}(R)}$ (with $L>0$)
so the integrand becomes just a function of $R$. Also in the SdS
case 
\begin{equation}
A_{O}(R)=-\frac{H_{O}^{2}}{R}\left(R-R_{-}\right)\left(R-R_0^{b}\right)\left(R-R_0^{c}\right)\label{eq:AO}
\end{equation}
where $R_{-}<0,\,R_0^{b}<R_0^{c}$ and $A_{O}$ is positive for
$R_0^{b}<R<R_0^{c}$. Now working within one panel of the infinite
strip of Penrose panels the one which contains the wall, the parametrization
$R=R(r)$ is one-valued we have for the first integral in \eqref{eq:I1-I2}
\[
\int_{\hat{r}}^{\infty}dr\frac{1}{2}\left(\frac{d}{dr}R^{2}\theta(-R_{+}')\right)=\frac{1}{2}\int_{\hat{R}}^{R_{\infty}}d\left(R^{2}\theta(\sqrt{A_{O}(R}))\right)=\frac{1}{2}\left(R_0^{b2}-\hat{R}^{2}\right)\theta\left(-\hat{R}'_{+}\right)
\]
In the last step we've used the fact that $R$ must decrease as one
goes from the lower limit to the upper limit and that the integral
gets cutoff at the first zero of $A_{O}$ as $R$ decreases - i.e.
at $R_0^{b}$, and the step function is the statement that in this
regime the sign of $R'$ must match the sign of $\hat{R}'_{+}$.

The second term in \eqref{eq:I1-I2} gives a delta function and so
is disconnected from the wall. Let us evaluate it explicitly. 
\begin{align}
-\int_{\hat{r}}^{\infty}dr\frac{1}{2}R^{2}\frac{d}{dr}\theta(-R_{+}') & =-N_{O}\int_{\hat{R}}^{R_{\infty}}dR\frac{1}{2}R^{2}\frac{d}{dR}\theta\left(\sqrt{A_{O}\left(R\right)}\right)=-N_{O}\int_{\hat{R}}^{R_{\infty}}dR\frac{1}{2}R^{2}\frac{d\sqrt{A_{O}}}{dR}\delta\left(\sqrt{A_{O}}\right)\nonumber \\
 & =-N_{O}\int_{\hat{R}}^{R_{\infty}}dR\frac{1}{2}R^{2}\frac{dA_{O}}{dR}\sum_{i=s,c}\frac{1}{|A_{O}(R_{i})|}\delta\left(R-R_{i}\right)\nonumber \\
 & =-N_{O}\int_{\hat{R}}^{R_{\infty}}dR\frac{1}{2}R^{2}\sum_{i=s,c}\epsilon\left(\frac{dA_{O}}{dR}|_{R_0^{i}}\right)\delta\left(R-R_0^{i}\right).\label{eq:discO}
\end{align}
Here $N_{O}$ is the number of panels included in the $r$ integral.
Using \eqref{eq:AO} we have
\[
\frac{dA_{O}}{dR}|_{R_0^{c}}=-\frac{H_{O}^{2}}{R_0^{c}}\left(R_0^{c}-R_0^{b}\right)<0,\,\,\frac{dA_{O}}{dR}|_{R_0^{b}}=-\frac{H_{O}^{2}}{R_0^{c}}\left(R_0^{b}-R_0^{c}\right)>0.
\]
Hence we have
\begin{align*}
-\int_{\hat{r}}^{\infty}dr\frac{1}{2}R^{2}\frac{d}{dr}\theta(-R_{+}') & =-N_{O}\sum_{i=b,c}\epsilon\left(\frac{dA_{O}}{dR}|_{R_0^{i}}\right)\frac{1}{2}R_0^{i2}\left\{ \theta\left(R_{\infty}-\hat{R}\right)\theta\left(R_{h}^{i}-\hat{R}\right)+\theta\left(\hat{R}-R_{\infty}\right)\theta\left(R_{h}^{i}-\hat{R}\right)\right\} \\
 & =N_{O}\left[R_0^{c2}\theta\left(R_0^{c}-\hat{R}\right)-R_0^{b2}\theta\left(R_0^{b}-\hat{R}\right)\right]\theta\left(R_{\infty}-\hat{R}\right)\\
 & +N_{O}\left[R_0^{c2}\theta\left(\hat{R}-R_0^{c}\right)-R_0^{b2}\theta\left(\hat{R}-R_0^{s}\right)\right]\theta\left(\hat{R}-R_{\infty}\right)
\end{align*}
For $\hat{R}=R_{2}$, since $R_0^{b}<R_{2}<R_0^{c}$ the above
gives
\[
N_{O}\left\{ \left(R_0^{c2}-0\right)\theta\left(R_{\infty}-\hat{R}\right)+\left(0-R_0^{b2}\right)\theta\left(\hat{R}-R_{\infty}\right)\right\} ,
\]
while for $\hat{R}=R_{1}$, since $R_0^{b}<R_{1}<R_0^{c}$ we
get exactly the same result since $N_{O}$ is independent of $\hat{R}$.
Thus when subtracting the two we get zero. In other words the disconnected
contribution cancels between the two turning points.

\subsection{dS horizon issues}

Here we explain using the explicit parametrization used in  \cite{DeAlwis:2019rxg} how the last term in eqn.\eqref{eq:SBHH}
arises from including a term corresponding to the nucleation of the brane behind the horizon of the observer.

In that calculation, at the turning point geometry\footnote{As observed in footnote \eqref{foot:explicit}.} we had
$R=H_{I}^{-1}\sin r_{-}$ (with $L=H_{I}^{-1}$) inside and $R=H_{O}^{-1}\sin r_{+}$
(with $L=H_{O}^{-1}$) outside. Since the physical radius $R$ must
be continous at the wall $R=\hat{R}=H_{I}^{-1}\sin\hat{r}_{-}=H_{O}\sin\hat{r}_{+}$.
Also $\frac{\ensuremath{\hat{R}_{\pm}'}}{L}=\cos\hat{r}_{\pm}$and
$A_{O/I}=1-H_{O/I}^{2}R^{2}=\cos^{2}r_{\pm}$. Define the static patch
identified as region III of the Penrose diagram as the one covered
by the range $0\le r_{\pm}<\pi/2$ and region I as the one covered
by $\pi/2<r_{\pm}<\pi$ . Note that at $r=\pi/2$ the static patch
coordinates (which have the factor $(1-H^{2}R^{2})$) become singular.

On the other hand $r$ is a global coordinate which may be extended
beyond $\pi/2$. But going beyond this point is tantamount to going
into region III from I or the reverese, in the Penrose diagram. In
fact since there are two values of $r$ determining a given $R$ (since
$\sin r=\sin(\pi-r)$) - one corresponding to region I and one for
region III.

So if we confine the calculation to region I, the outside integral
(the second term in \eqref{eq:SButp}) becomes after putting $\sin r=s$,
(note that $RR'=H_{O}^{-2}\sin r\cos r$)
\begin{align}
-\frac{\eta}{G}H_{O}^{-2}\int_{\hat{r}_{+}}^{\pi}dr\sin r\cos r\cos^{-1}\left(\frac{\cos r}{|\cos r|}\right) & =-\frac{\eta}{G}H_{O}^{-2}\int_{\sin\hat{r}_{+}}^{0}dss\pi\theta\left(\sin^{-1}s-\frac{\pi}{2}\right)\nonumber \\
 & =\frac{\eta\pi}{2G}H_{O}^{-2}\sin^{2}\hat{r}_{+}\theta\left(\hat{r}_{+}-\frac{\pi}{2}\right)=\frac{\eta\pi}{2G}\hat{R}^{2}\theta\left(-\hat{R}'_{+}\right).\label{eq:dSout}
\end{align}
For the inside integral (the first integral in \eqref{eq:SButp})
we have 
\begin{align}
-\frac{\eta}{G}H_{I}^{-2}\int_{0}^{\hat{r}_{-}}dr\sin r\cos r\cos^{-1}\left(\frac{\cos r}{|\cos r|}\right) & =-\frac{\eta\pi}{G}H_{I}^{-2}\int_{0}^{\sin\hat{r}_{-}}dss\theta\left(\sin^{-1}s-\frac{\pi}{2}\right)\nonumber \\
 & =-\frac{\eta\pi}{2G}H_{I}^{-2}\left[\sin^{2}\hat{r}-\sin^{2}\frac{\pi}{2}\right]\theta\left(\sin^{-1}\hat{s}-\frac{\pi}{2}\right)\nonumber \\
 & =-\frac{\eta\pi}{2G}\left[\hat{R}^{2}-H_{I}^{-2}\right]\theta\left(-\hat{R}'_{-}\right).\label{eq:dSin}
\end{align}
Thus for the total bulk action we get
\begin{equation}
I\left(\hat{R}\right)=iS\left(\hat{R}\right)=\frac{\eta\pi}{2G}\left[\left(\theta\left(-\hat{R}'\right)-\theta\left(-\hat{R}'_{-}\right)\right)\hat{R}^{2}+H_{I}^{-2}\theta\left(-\hat{R}'_{-}\right)\right]\label{eq:dSdSStaticPatch}
\end{equation}
This expression is exactly what is obtained in the the limit $R_{O,b}\rightarrow0$
of the SdS to SdS expression (i.e. \eqref{eq:SdStoSdS} for the bulk integral) - namely eqn. \eqref{eq:SBHH1}.

Comparing this expression to eqn. \eqref{eq:SBHH}, we see that it
is missing the term $\theta(\hat{R}'_{+})H_{O}^{2}$ within the square
brackets. Where did this come from? 

In the calculation in \citep{DeAlwis:2019rxg} we used global coordinates
to calculate the bulk integral. In terms of the static patch coordinates
relevant to an observer in a given such patch - this involves integration
through the deSitter horizon. To see this in the above calculation
we need to add a piece proportional to $\theta\left(\frac{\pi}{2}-\hat{r}\right)$
corresponding to the brane being nucleated behind the horizon of the
observer at $r=\pi$. Thus we add the term,
\begin{equation}
-\frac{\eta}{G}H_{O}^{-2}\int_{\sin\hat{r}_{+}}^{\sin\pi}dss\pi\theta\left(\sin^{-1}s-\frac{\pi}{2}\right)\theta\left(\frac{\pi}{2}-\hat{r}\right)=-\frac{\eta\pi}{G}H_{O}^{-2}\int_{\sin\frac{\pi}{2}}^{\sin\pi}dss\theta\left(\frac{\pi}{2}-\hat{r}\right)=\frac{\eta\pi}{2G}H_{O}^{-2}\theta\left(\hat{R}'_{+}\right).\label{eq:dSOadd}
\end{equation}
Adding this to \eqref{eq:dSdSStaticPatch} gives the result \eqref{eq:SBHH}

It should be noted that we could have done the integral in \eqref{eq:dSout}
in a differnt way by writing (after identifying the intgration variable
as $t=\cos r$)

\begin{align*}
-\frac{\eta}{G}H_{O}^{-2}\int_{\hat{r}_{+}}^{\pi}dr\sin r\cos r\cos^{-1}\left(\frac{\cos r}{|\cos r|}\right) & =\frac{\eta}{G}H_{O}^{-2}\int_{\cos\hat{r}_{+}}^{\cos\pi}dtt\pi\theta\left(-t\right)=\frac{\eta\pi}{2G}H_{O}^{-2}\left[1-\left(1-\sin^{2}\hat{r}_{+}\right)\theta\left(-\cos\hat{r}_{+}\right)\right]\\
 & =\frac{\eta\pi}{2G}\left[H_{O}^{-2}\theta\left(\hat{R}_{+}\right)+\hat{R}^{2}\theta\left(-\hat{R}'_{+}\right)\right].
\end{align*}
However this way of calculating (essentially what was done in \citep{DeAlwis:2019rxg})
misses the fact that the extra term comes from going behind the horizon
of the observer at $r=\pi$ . 

To belabor the point let us redo the calculation without using an
explicit parametrization. This is all we can do in the general situation
of SdS transitions where we do not have the luxury of such a parametrization.
In this case the outside integral (the second term in \eqref{eq:SButp})
may written as (ignoring the delta function terms discussed in the previous subsection)
\begin{align*}
 & =-\frac{\eta\pi}{G}\left[\int_{\hat{r}}^{r_{\rm max}}drR'R\theta(-R_{+}')\right]=-\frac{\eta\pi}{2G}\left[\int_{\hat{r}}^{r_{\rm max}}dr\frac{dR^{2}}{dr}\theta(-R_{+}')\right]=-\frac{\eta\pi}{2G}\left[\int_{\hat{R}}^{0}dR^{2}\theta(-R_{+}')\right]\\
 & =\frac{\eta\pi}{2G}\hat{R}^{2}\theta(-R_{+}')
\end{align*}
In agreement with \eqref{eq:dSout}.

\subsection{Reflection symmetry of tunneling action}\label{subsec:Reflection-symmetry-of}
In this subsection we will show explicitly the symmetry under the interchange of the outside and inside space times when the configuration is at a general point (ii.e. not a turning point), for both the bulk and the brane actions.

Let us first look at ($i$ times) the bulk action $I=iS$, with the
wall/brane at some arbitrary point with radius $\hat{R}=R(\hat{r})$.
We have for a transition from a state $O$ outside to a state $I$
inside\footnote{Note that we've replaced $r_{\rm max}$ by $r*$.},
\begin{align}
\frac{G}{\eta}I_{B}^{IO} & =\int_{0}^{\hat{r}-\epsilon}drR\left[\sqrt{A_{I}L^{2}-R'^{2}}-R'\cos^{-1}\left(\frac{R'}{LA_{I}}\right)\right]\nonumber \\
 & +\int_{\hat r+\epsilon}^{r*}drR\left[\sqrt{A_{O}L^{2}-R'^{2}}-R'\cos^{-1}\left(\frac{R'}{LA_{O}}\right)\right]\label{eq:IO}
\end{align}

\begin{align}
\frac{G}{\eta}I_{B}^{OI} & =\int_{0}^{\hat{r}}drR\left[\sqrt{A_{I}L^{2}-R'^{2}}-R'\cos^{-1}\left(\frac{R'}{LA_{I}}\right)\right]\nonumber \\
 & +\int_{\hat{r}}^{r*}drR\left[\sqrt{A_{O}L^{2}-R'^{2}}-R'\cos^{-1}\left(\frac{R'}{LA_{O}}\right)\right]\label{eq:OI}
\end{align}
In the first equation let us change the integration variable: put
$r'=r*-r$. So we get 

\begin{align}
\frac{G}{\eta}I_{B}^{IO} & =\int_{r*}^{\hat{r}'+\epsilon}\left(-dr'\right)R\left(r*-r'\right)\left[\sqrt{A_{I}L^{2}-R'^{2}}-\frac{dR}{d\left(r*-r'\right)}\cos^{-1}\left(\frac{dR/d\left(r*-r'\right)}{LA_{I}}\right)\right]\nonumber \\
 & +\int_{\hat{r}-\epsilon}^{0}\left(-dr'\right)R\left(r*-r'\right)\left[\sqrt{A_{O}L^{2}-R'^{2}}-\frac{dR}{d\left(r*-r'\right)}\cos^{-1}\left(\frac{dR/d\left(r*-r'\right)}{LA_{O}}\right)\right]\label{eq:IO-1}
\end{align}
The above may be rewritten as
\begin{align*}
\frac{G}{\eta}I_{B}^{IO}\left[R\left(\hat{r}\right)\right] & =\int_{\hat{r}'+\epsilon}^{r*}\left(-dr'\right)R\left(r*-r'\right)\left[\sqrt{A_{I}L^{2}-R'^{2}}-\frac{dR}{dr}\vert_{r=r*-r'}\cos^{-1}\left(\frac{\left(dR/dr\right)\vert_{r=r*-r'}}{LA_{I}}\right)\right]\\
 & +\int_{0}^{\hat{r}-\epsilon}drR\left(r*-r'\right)\left[\sqrt{A_{O}L^{2}-R'^{2}}-\frac{dR}{dr}\vert_{r=r*-r'}\cos^{-1}\left(\frac{\left(dR/dr\right)\vert_{r=r*-r'}}{LA_{O}}\right)\right]\\
 & =\text{\ensuremath{\frac{G}{\eta}I_{B}^{OI}\left[R\left(r*-\hat{r}'\right)\right]}=\ensuremath{\frac{G}{\eta}I_{B}^{OI}\left[R\left(\hat{r}\right)\right]}},
\end{align*}
which establishes the symmetry of the bulk action between the outside
and inside space-times.

For the wall/brane action we have, 
\begin{equation}
-\frac{G}{\eta}I_{W}^{IO}=\int\delta\hat{R}\hat{R}\left[\cos^{-1}\left(\frac{\hat{R}_{+}}{L\sqrt{A_{O}}}\right)-\cos^{-1}\left(\frac{\hat{R}_{-}}{L\sqrt{A_{I}}}\right)\right]\label{eq:WIO}
\end{equation}

Under the interchange $O\rightleftarrows I$ we have $\hat{R}_{\pm}\rightleftarrows-\hat{R}_{\mp}$.
So since $\cos^{-1}\left(-x\right)=\pi-\cos^{-1}\left(x\right)$,
we have 
\begin{align*}
-\frac{G}{\eta}I_{W}^{OI} & =\int\delta\hat{R}\hat{R}\left[\left(\pi-\cos^{-1}\left(\frac{\hat{R}_{-}}{L\sqrt{A_{I}}}\right)\right)-\left(\pi-\cos^{-1}\left(\frac{\hat{R}_{+}}{L\sqrt{A_{O}}}\right)\right)\right]\\
 & =\int\delta\hat{R}\hat{R}\left[\left(-\cos^{-1}\left(\frac{\hat{R}_{-}}{L\sqrt{A_{I}}}\right)\right)+\left(\cos^{-1}\left(\frac{\hat{R}_{+}}{L\sqrt{A_{O}}}\right)\right)\right]=-\frac{G}{\eta}I_{W}^{IO}.
\end{align*}
Hence the total action at $\hat{R}$ is invariant under the interchange
of the outside and the inside.

It is important to note that this reflection symmetry is manifest only with at general points in field space - i.e. without imposing turning point values for either the geometry or
the brane.

\bibliographystyle{utphys}
\bibliography{references.bib}

\end{document}